\documentclass[a4j, 11pt,autodetect-engine
]{article}
\usepackage{jheppub}

\usepackage{wrapfig}
\usepackage[utf8]{inputenc}
\usepackage{braket}

\hypersetup{
    colorlinks=true,
    linkcolor=blue,
    citecolor=cyan,
    urlcolor=blue,
    pdfpagemode=FullScreen,
    }
    
\usepackage{amsmath,amsthm,amsfonts,amssymb,amscd}
\usepackage{bm}
\usepackage{multirow,booktabs}
\usepackage[table,svgnames]{xcolor}
\usepackage{enumerate}
\usepackage{enumitem}
\usepackage{mathrsfs}
\usepackage{wrapfig}
\usepackage{slashed}

\edef\restoreparindent{\parindent=\the\parindent\relax}
\usepackage{parskip}
\restoreparindent

\usepackage{tikz}
\usepackage{tikz-3dplot}
\usetikzlibrary{arrows,arrows.meta,intersections, calc,positioning,decorations.pathreplacing,decorations.pathmorphing,shapes}
\usetikzlibrary{patterns}
\usetikzlibrary{decorations.markings}

\def\i{{\rm i}}
\def\d{{\rm d}}
\def\e{{\rm e}}
\def\tr{{\rm tr}}
\def\CC{\mathcal{C}}
\def\CD{\mathcal{D}}

\def\CK{\mathcal{K}}

\def\CM{\mathcal{M}}

\def\CO{\mathcal{O}}
\def\CR{\mathcal{R}}

\def\BH{\mathbb{H}}

\def\BR{\mathbb{R}}
\def\BS{\mathbb{S}}
\def\BZ{\mathbb{Z}}

\def\csch{\mathrm{csch}}

\def\SO{\mathrm{SO}}

\vspace*{0cm}

\title{Holographic defect CFTs with Dirichlet end-of-the-world branes}
\author{Haruki Nakayama}
\author{and Tatsuma Nishioka}

\affiliation{Department of Physics, The University of Osaka,\\
Machikaneyama-Cho 1-1, Toyonaka 560-0043, Japan}

\preprint{OU-HET-1291}

\abstract{We construct a holographic model of defect conformal field theories (DCFTs) with defects of codimension greater than one.
Our construction generalizes the AdS/BCFT model by anchoring the end-of-the-world brane on defects at the asymptotic AdS boundary and imposing Dirichlet boundary conditions for the metric on the brane.
We compute the defect entropy and defect free energy and show that the defect $\CC$-function is always non-negative.
We further study holographic defect-localized RG flows triggered by a localized scalar field on the brane and show that the defect $\CC$-theorem holds.
We also verify that our model reproduces the expected forms of correlation functions in DCFTs.
}

\begin{document}

\maketitle

\section{Introduction}

Understanding non-perturbative aspects of quantum field theories (QFTs) is a central challenge in modern theoretical physics.
Conventional methods based on perturbation theory fail in strongly coupled regimes, except in rare cases with additional structures such as integrability or supersymmetry. 
While it is generally impossible to track the dynamics of QFTs along renormalization group (RG) flows, the theories are more accessible at critical points of RG flows where conformal symmetry emerges and conformal field theories (CFTs) provide invaluable tools to place strong constraints on correlation functions and the spectrum of operators \cite{Belavin:1984vu,DiFrancesco:1997nk}.

In the real world, physical systems are often of finite size or involve interfaces and impurities.
They take various forms, such as lines (e.g., Wilson-'t Hooft lines, magnetic impurities or cosmic strings), domain walls or boundaries, and termed defects collectively. 
In string theory, defects
are represented by D-branes \cite{Polchinski:1995mt,Polchinski:1998rq}, which extend into higher dimensions. 
Topological defects also play distinguished roles as symmetry generators and charged objects in generalized symmetries \cite{Gaiotto:2014kfa}.

The presence of defects typically breaks the whole symmetry even at the critical points of the systems.
However, a part of conformal symmetry can be preserved when defects are either planar or spherical.
Such defects are known as conformal defects, and 
CFTs that incorporate them are referred to as defect conformal field theories (DCFTs).
More concretely, a conformal defect $\CD^{(p)}$ of dimension $p$ breaks the ambient (Euclidean) conformal symmetry group $\SO(1,d+1)$ in $d$-dimensions to the subgroup $\SO(1,p+1) \times \SO(d-p)$, which act as the conformal symmetry on $\CD^{(p)}$ and the rotational symmetry around it, respectively.
This residual symmetry is still stringent enough to constrain correlation functions and operator spectrum in DCFTs \cite{McAvity:1995zd,Liendo:2012hy,Billo:2016cpy,Gadde:2016fbj,Basu:2015gpa,Nii:2016lpa,Lemos:2017vnx,Isachenkov:2017qgn,Soderberg:2017oaa,Kobayashi:2018okw,Guha:2018snh,Lauria:2018klo,Liendo:2019jpu,Herzog:2020bqw,Giombi:2020rmc,Dey:2020jlc,Giombi:2021cnr,Nishioka:2022ook,Sato:2021eqo,Giombi:2021uae,Bianchi:2021snj,Collier:2021ngi,Giombi:2022vnz,Nishioka:2022odm,Nishioka:2022qmj,Bissi:2022bgu,Gimenez-Grau:2022czc,Gimenez-Grau:2022ebb,Bianchi:2022sbz,Dey:2024ilw}.

Besides correlation functions, there are global observables that probe localized degrees of freedom on defects such as defect entropy and defect free energy \cite{Affleck:1991tk,Calabrese:2004eu,Azeyanagi:2007qj,Nozaki:2012qd,Jensen:2013lxa,Estes:2014hka,Gaiotto:2014gha,Herzog:2015ioa,Jensen:2015swa,Estes:2018tnu,Kobayashi:2018lil,Jensen:2018rxu,Chalabi:2020iie,Nishioka:2021uef,Chalabi:2021jud,Chalabi:2022qit,Yuan:2022oeo,Yuan:2023oni,Capuozzo:2023fll,Capuozzo:2024onf,Apruzzi:2024ark,Ge:2024hei}.
They provide a natural candidate for defect $\CC$-functions, which decrease monotonically under defect-localized RG flows \cite{Nozaki:2012qd,Estes:2014hka,Gaiotto:2014gha,Herzog:2015ioa,Jensen:2015swa,Kobayashi:2018lil,Casini:2016fgb,Wang:2020xkc,Cuomo:2021rkm,Wang:2021mdq}.
Field theory results, including free and weakly coupled theories, support these expectations, though general proofs remain elusive except in special cases \cite{Affleck:1991tk,Jensen:2015swa,Casini:2016fgb,Cuomo:2021rkm,Wang:2021mdq,Casini:2022bsu,Casini:2018nym,Casini:2023kyj,Harper:2024aku}.
On the other hand, holography serves as an effective approach to studying such observables by recasting their computation as geometric problems in the dual gravitational description.

Holographic dualities for DCFTs have been investigated as natural generalizations of the AdS/CFT correspondence.
In string and supergravity theories, many exact solutions with smooth geometries dual to DCFTs have been constructed in e.g.\,\cite{DeWolfe:2001pq,Bachas:2001vj,Bak:2003jk,DHoker:2006qeo,Hirano:2006as,Yamaguchi:2006te,Lunin:2006xr,Lunin:2007ab,DHoker:2007zhm,DHoker:2007hhe,Gutperle:2019dqf,Gutperle:2020gez,Gutperle:2020rty,Chen:2019qib,Chen:2020mtv}.
In bottom-up models, pure tension branes are introduced as probes dual to defects \cite{Karch:2000gx,Yamaguchi:2002pa,Chang:2013mca,Jensen:2013lxa,Kobayashi:2018lil} or as boundaries that terminate the bulk AdS space so as to describe boundary CFTs (BCFTs) \cite{Takayanagi:2011zk,Fujita:2011fp,Miyaji:2022dna}.
Despite these advances, the existing holographic models of DCFTs are limited to describing defects of specific types and dimensions or treating them as probes.
Hence, it is desirable to construct holographic models capable of describing conformal defects of diverse dimensions in a unified manner while reproducing the expected structural features of DCFTs.

In this paper, we address this gap by proposing a bottom-up holographic model of DCFTs that accommodate defects of arbitrary dimensions.
Our construction stands on the viewpoint that a defect $\CD^{(p)}$ can be characterized by boundary conditions imposed on a small tubular neighborhood around $\CD^{(p)}$, even for $p< d-1$.
We realize the holographic dual of such DCFTs as a limit of the AdS/BCFT \cite{Takayanagi:2011zk,Fujita:2011fp}, where the EoW brane is anchored on the tubular neighborhood at the asymptotic AdS boundary.
While Neumann boundary conditions are imposed on the brane in the AdS/BCFT model, in our higher-codimensional setup these conditions yield unphysical solutions in which the bulk AdS space collapses completely.
To resolve this issue, we employ Dirichlet boundary conditions instead, which lead to physical solutions with finite bulk regions and positive brane tensions.

We test the validity of our model through several holographic computations.
First, we evaluate defect entropy and defect free energy holographically.
In field theories with a UV cutoff $\epsilon$, they are known to have the UV structures of order $O(1/\epsilon^{p-2})$ and $O(1/\epsilon^{p})$ for a $p$-dimensional defect, respectively.
We show our holographic calculations reproduce these structures correctly.
We then calculate the defect $\CC$-function in our model and show that it is always non-negative for physical configurations of the EoW brane.
Furthermore, we model holographic defect-localized RG flows by introducing a localized scalar field on the brane and prove the defect $\CC$-theorem by showing the defect $\CC$-function decreases under such flows.
We also examine the correlation functions of scalar primaries with two methods: by coupling a bulk scalar field to the EoW brane, and the geodesic approximation.
In both cases, our model reproduces the expected forms of the correlation functions, further supporting its consistency as a holographic description of DCFTs with defects of arbitrary dimensions.

The rest of the paper is organized as follows.
In section \ref{sec:DCFT_review}, we review relevant aspects of DCFTs, including their symmetry structures, correlation functions, and the notions of defect entropy and defect free energy, as well as the current status of the $\CC$-theorem in DCFTs.
Section \ref{sec:Hol_DCFT} introduces our holographic construction of DCFTs, beginning with a review of the AdS/BCFT model, and then extending it to the AdS/DCFT model with Dirichlet boundary conditions on the EoW brane.
We compute defect entropies, defect free energies, and defect $\CC$-functions in this setup.
Then, we present a holographic proof of the defect $\CC$-theorem for defect-localized RG flows triggered by a brane-localized scalar field.
In section \ref{sec:Hol_correlators}, we examine correlation functions in our holographic model by two methods and reproduce the expected forms from DCFTs.
Finally, section \ref{sec:discussion} concludes with a discussion and potential directions for future works.

\section{Defect CFTs}\label{sec:DCFT_review}
We begin with reviewing the key results in DCFTs that will be relevant in describing our holographic model in later sections.
In section \ref{ss:coordinates}, we introduce the coordinate systems that make the $\SO(1,p+1) \times \SO(d-p)$ symmetry manifest in a DCFT with a $p$-dimensional conformal defect $\CD^{(p)}$.
Section \ref{ss:correlation_functions} summarizes the structures of one- and two-point functions in DCFTs, and section \ref{subsec:entropy_free_energy} introduces the notions of defect entropy and defect free energy.
Finally, section \ref{ss:defect_C} provides an overview of the current status of the $\CC$-theorem in DCFTs.

\subsection{Conformal defects and coordinate transformation}\label{ss:coordinates}

\begin{figure}[t]
    \centering
    \begin{tikzpicture}[scale=1.0,>=stealth]

    \fill[orange!30] (-0.5,0) -- (-0.5,2) -- (0.5,2.5) -- (0.5,0.5) -- cycle;
    \draw (-0.5,0) -- (-0.5,2) -- (0.5,2.5) -- (0.5,0.5) -- cycle;
    \node at (-1.0,1.25) {$\CD^{(p)}$};
    \node at (0.0,-0.5) {$\BR^d$};
    
    \draw[->] (1.25,1.25) -- (4.25,1.25) node[midway,above] {conformal map};

    \begin{scope}[shift={(6.0,1.25)}]
        \draw (2.0,0.2) parabola (0.04,0.99);
        \draw (2.0,-0.2) parabola (0.04,-0.99);
        \draw (2.0,0.2) arc (90:-90:0.1 and 0.2);
        \fill[orange!30] (0,0) circle [x radius = 0.3, y radius = 1.0];
        \draw (0,0) circle [x radius = 0.3, y radius = 1.0];
        \node at (-0.8,0.0) {$\CD^{(p)}$};
        \node at (1.0,-1.75) {$\BH^{p+1}$};
    \end{scope}

    \node at (9.0,1.25) {\Large $\times$};
    
    \begin{scope}[shift={(11.0,1.25)}]
        \draw (0,0) circle (1.0);
        \draw (-1,0) arc (180:360:1 and 0.35);
        \draw[dashed] (-1,0) arc (180:0:1 and 0.35);
        \node at (0,-1.75) {$\BS^{d-p-1}$};
    \end{scope}
    
    \end{tikzpicture}
    \caption{A $p$-dimensional planar defect $\CD^{(p)}$ in $\BR^{d}$ is mapped to the boundary of  the hyperbolic space in $\BH^{p+1} \times \BS^{d-p-1}$ under the conformal transformation.}
    \label{fig:conformal_map}
\end{figure}
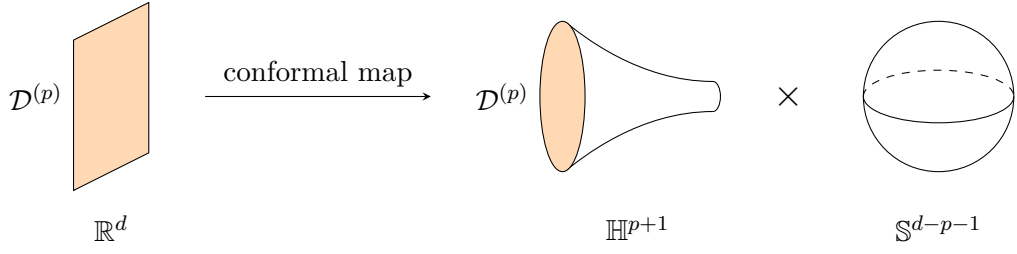

We consider a $p$-dimensional planar conformal defect $\CD^{(p)}$ in a CFT on $\BR^d$.
Let $\hat x^{\hat a}~(\hat a = 0, \cdots ,\,p-1)$ and $x_\perp^i~ (i = p, \cdots ,\,d-1)$ be the parallel and transverse components of the $d$-dimensional coordinates $x^\mu = (\hat x^{\hat a},\,x_\perp^i)$ to $\CD^{(p)}$:
\begin{align}\label{planar_defect}
    \CD^{(p)} 
        = \left\{\,  x^\mu \in \BR^d~ \Big|~x_\perp^{p} = \cdots =  x_\perp^{d-1} = 0\, \right\} \ .
\end{align}
The metric of $\BR^d$ is divided into the parallel and transverse parts as follows:
\begin{align}\label{flat_metric}
    \begin{aligned}
        \d s^2 
            &= 
                \d \hat{x}^{\hat a}\, \d \hat{x}^{\hat a} + \d x_\perp ^i\, \d x_\perp ^i \\
            &=:
                \d \hat{x}^2_{\hat{a}} + \d x^2_{\perp,i}
                \ .    
    \end{aligned}
\end{align}
To make manifest the residual conformal symmetry $\SO(1,p+1) \times \SO(d-p)$ in the presence of conformal defects, we use the polar coordinate $\d x^2_{\perp,i} = \d r^2 + r^2\,\d \Omega_{d-p-1}^2$ for the transverse coordinates and rewrite the metric \eqref{flat_metric} as
\begin{align}\label{polar_coordinate}
    \begin{aligned}
        \d s^2 
            &=
                \d \hat{x}^2_{\hat{a}} + \d r^2 + r^2\,\d \Omega_{d-p-1}^2
                \\
            &=
                r^2\left(\frac{\d \hat{x}^2_{\hat{a}} + \d r^2}{r^2} + \d \Omega_{d-p-1}^2\right)
                \ ,   
    \end{aligned}  
\end{align}
where $\d \Omega_{d-p-1}^2$ is the metric for a unit $(d-p-1)$-dimensional sphere $\BS^{d-p-1}$.
We thus find that the flat space is conformally equivalent to $\BH^{p+1}\times\BS^{d-p-1}$ with the metric:
\begin{align}\label{hyperbolic_coordinate}
    \d s^2_{\BH^{p+1}\times\BS^{d-p-1}}
        =
            \frac{\d \hat{x}^2_{\hat{a}} + \d r^2}{r^2} + \d \Omega_{d-p-1}^2 \ .
\end{align}
In the new coordinate, the residual conformal symmetry $\SO(1,p+1) \times \SO(d-p)$ is realized as the isometry of $\BH^{p+1}\times \BS^{d-p-1}$.
The conformal defect $\CD^{(p)}$ is located at the boundary of the $(p+1)$-dimensional
hyperbolic space $\BH^{p+1}$ (see figure \ref{fig:conformal_map}):
\begin{align}
    \CD^{(p)} 
        = \left\{\,x^\mu \in \BR^{d}~ \Big|~r=0\,\right\} \ .
\end{align}

\begin{figure}[t]
    \centering
    \begin{tikzpicture}[scale=1.0, >=stealth]

    \draw[dashed] (-0.8,0) arc (180:0:0.8 and 0.3);
    \draw[dashed] (-0.8,3.0) arc (180:0:0.8 and 0.3);
    \draw[thick,orange] (0,-0.5) -- (0,3.5);
    \node at (0,3.8) {$\CD^{(p)}$};
    \draw (-0.8,0) arc (180:360:0.8 and 0.3);
    \draw (-0.8,3.0) arc (180:360:0.8 and 0.3);
    \draw (0.8,0) -- (0.8,3.0);
    \draw (-0.8,0) -- (-0.8,3.0);
    \draw[{Stealth[angle'=60,scale=0.8]}-{Stealth[angle'=60,scale=0.8]}] (0,0.0) -- (0.8,0.0);
    \draw[{Stealth[angle'=60,scale=0.8]}-] (0.85,1.6) arc [start angle = 30, end angle = -210, x radius = 1.0, y radius = 0.3];
    \draw[{Stealth[angle'=60,scale=0.8]}-] (-1.1,2.5) -- (-1.1,0.5);
    \draw (1.5,4.5) -- (1.5,4.0) -- (2.0,4.0);
    \node at (1.8,1.6) {$\BS^{d-p-1}$};
    \node at (-1.4,2.0) {$\BR^{p}$};
    \node at (1.2,2.5) {$N_{\epsilon}$};
    \node at (0.4,0.12) {$\epsilon$};
    \node at (1.8,4.3) {$\BR^{d}$};
    \fill[gray!30, opacity=0.5] (-0.8,0) arc (180:360:0.8 and 0.3) -- (0.8,3.0) arc (360:180:0.8 and 0.3) -- cycle;
    
    \end{tikzpicture}
    \caption{The codimension-one surface $N_{\epsilon}$ of the tubular neighborhood of a defect $\CD^{(p)}$.}
    \label{fig:tubular}
\end{figure}
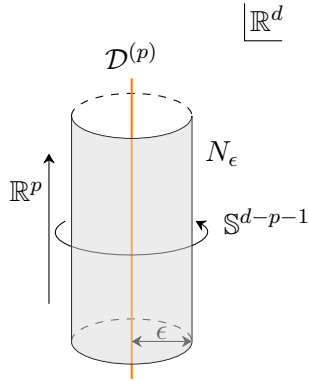

One may view $\CD^{(p)}$ as a boundary condition for fields in the ambient CFT on $\BH^{p+1}\times \BS^{d-p-1}$ which preserves the full isometry of the background \cite{Kapustin:2005py,Herzog:2020lel,Giombi:2020rmc}.
The situation is similar to the AdS/CFT correspondence, motivating us to introduce the cut-off surface at $r=\epsilon~(\ge 0)$ where the boundary condition characterizing the defect is imposed.
In the flat space metric \eqref{polar_coordinate}, introducing such a cut-off surface corresponds to excising the tubular neighborhood around the defect with the boundary surface $N_\epsilon \simeq \BR^{p} \times \BS^{d-p-1}$ (see figure \ref{fig:tubular}):
\begin{align}
    N_\epsilon = \left\{\,  x^\mu \in \BR^d~ \Big|~r= \epsilon\,\right\} \ .
\end{align}
Note that $N_\epsilon$ is a codimension-one hypersurface, regardless of the dimension $p$ of a conformal defect  $\CD^{(p)}$.
This seemingly simple observation is a key step in our approach as it allows us to describe $\CD^{(p)}$ in terms of BCFTs on the manifold 
\begin{align}\label{M_DCFT}
    M = \left\{\, x^\mu \in \BR^d~ \Big|~ r\ge\epsilon\,\right\} \ ,    
\end{align}
whose boundary is $\partial M = N_\epsilon$.
We will exploit this description of conformal defects when we construct a holographic model of DCFTs in section \ref{sec:Hol_DCFT}.

\subsection{Correlation functions}\label{ss:correlation_functions}

In contrast to CFTs without defects, the one-point functions of the ambient operators in DCFTs do not vanish due to the lack of the translational symmetry transverse to the defect while the one-point functions of the defect localized operators still vanish as in the $p$-dimensional CFTs.
Despite the lack of the full conformal symmetry, the residual conformal symmetry can fix the form of the one-point functions of the ambient operators. 
On the other hand, the two-point functions exhibit a richer structure, depending on the cross-ratios that are invariant under the residual symmetry (see e.g., \cite{Billo:2016cpy,Gadde:2016fbj,Kobayashi:2018okw,Guha:2018snh,Lauria:2018klo}).
In what follows, we focus on the cases with scalar operator and stress tensor that will be needed in later sections.

Let us first consider the scalar correlation functions.
By inspecting the translational symmetry along the defect and scaling symmetry, the one-point function of the ambient scalar operator $\CO(x)$ is determined as
\begin{align}\label{one_point_scalar}
    \langle\,\CO(x)\,\rangle
        &=
            \frac{a_{\CO}}{|x_{\perp}|^{\Delta}} \ ,
\end{align}
where $a_{\CO}$ is a constant that cannot be fixed by symmetry considerations and $\Delta$ is the conformal dimension of $\CO(x)$.
For a pair of operators located at $x = x_1$ and $x_2$, one can construct the cross-ratios 
\begin{align}\label{cross-ratios}
    \xi_1
        = 
            \frac{|x_1-x_2|^2}{|x_{1,\perp}|\,|x_{2,\perp}|}
    \ , \qquad
    \xi_2
        =
            \frac{x_{1,\perp}\cdot\,x_{2,\perp}}{|x_{1,\perp}|\,|x_{2,\perp}|}
    \ ,
\end{align}
invariant under the residual conformal symmetry.
The scalar two-point function is fixed up to a scalar function $f(\xi_1,\xi_2)$ of the cross-ratios as
\begin{align}\label{two_point_scalar}
    \langle\,\CO_1(x_1)\,\CO_2(x_2)\,\rangle
        &=
            \frac{f(\xi_1,\xi_2)}{|x_{1,\perp}|^{\Delta_1}\,|x_{2,\perp}|^{\Delta_2}} \ ,
\end{align}
where $\Delta_{1,2}$ are the conformal dimensions of the scalar operators $\CO_1(x_1),\CO_2(x_2)$ respectively.

Next, we turn to the one-point function of the ambient stress tensor.
The stress tensor is defined by
\begin{align}\label{stress_tensor}
    T^{\mu\nu}_{\text{DCFT}}
        =
            \frac{2}{\sqrt{g}}\frac{\delta\,\log Z_{\text{DCFT}}[g_{\mu\nu}]}{\delta\,g_{\mu\nu}} \ ,
\end{align}
where $Z_{\text{DCFT}}$ is the partition function of DCFT.\footnote
{Our definitions differs by its sign from the one in \cite{Kobayashi:2018lil}.}
The stress tensor can be split it into the ambient part $T^{\mu\nu}_{\text{CFT}}$ and the defect localized part $t^{\mu\nu}$ as
\begin{align}
    T^{\mu\nu}_{\text{DCFT}}
        =
            T^{\mu\nu}_{\text{CFT}} + t^{\mu\nu} \ .
\end{align}
While the one-point function of the defect localized stress tensor vanishes due to the conformal symmetry on the defect, the ambient part is fixed up to a constant $a_T$ by inspecting the residual conformal symmetry \cite{Kapustin:2005py,Billo:2016cpy}:\footnote
{Our $a_T$ is the same as the one used in \cite{Kobayashi:2018lil}. Note that our definition of the stress tensor \eqref{stress_tensor} differs from theirs by an an overall sign.}
\begin{align}
    \begin{aligned}\label{one_point_stress}
        \langle\,T^{\hat{a}\hat{b}}_{\text{CFT}}(x)\,\rangle
            &=
                - \frac{d-p-1}{d}\frac{a_T}{|x_{\perp}|^d}\,\delta^{\hat{a}\hat{b}} \ ,
        \\
        \langle\,T^{ij}_{\text{CFT}}(x)\,\rangle
            &=
                \frac{a_T}{|x_{\perp}|^d}\,\bigg(\frac{p+1}{d}\,\delta^{ij}-\frac{x_{\perp}^ix_{\perp}^j}{|x_{\perp}|^2}\bigg) \ ,
        \\
        \langle\,T^{\hat{a}i}_{\text{CFT}}(x)\,\rangle
            &=
                0 \ .
    \end{aligned}
\end{align}
Note that it automatically satisfies the conservation law $\partial_\mu \langle\,T^{\mu\nu}_{\text{CFT}}(x)\,\rangle = 0$ and $\langle\,T^{\mu\nu}_{\text{CFT}}(x)\,\rangle = 0$ for $p=d-1$, i.e., in BCFTs or interface CFTs (ICFTs).

\subsection{Defect entropy and defect free energy}\label{subsec:entropy_free_energy}
In this subsection, we introduce the notions of defect entropy and defect free energy, defined respectively as the excess of the entanglement entropy and sphere free energy induced by defects.

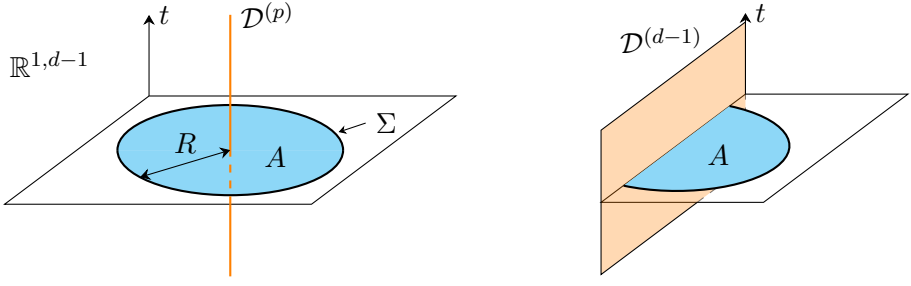
\begin{figure}[t]
\centering
  \begin{minipage}{0.35\hsize}
  \centering
  \begin{tikzpicture}[scale=1.2, >=stealth]
   \draw (0.5,0.4)--(3.9,0.4)--(5.5,1.6)--(2.1,1.6)--cycle;
   \draw[-{Stealth[angle'=60,scale=0.8]}] (2.1, 1.6)--(2.1, 2.5) node [right] {$t$};
   \draw[fill=cyan!40, thick] (4.25,1) arc [start angle = 0, end angle = 180, x radius=1.25cm, y radius=0.5cm];
   \draw[fill=cyan!40, thick] (4.25,1) arc [start angle = 0, end angle = -180, x radius=1.25cm, y radius=0.5cm];
   
   \draw[thick, orange] (3.0,1.0)--(3.0,2.5) node [black, right] {$\mathcal{D}^{(p)}$};
   \draw[dashed, thick, orange] (3.0,0.4)--(3.0,1.0);
   \draw[thick, orange] (3.0,-0.4)--(3.0,0.4);
   
   \draw[{Stealth[angle'=60,scale=0.8]}-{Stealth[angle'=60,scale=0.8]}] (2.0,0.7)--(2.5,0.85) node [above] {$R$} --(3.0,1.0);
   \draw[{Stealth[angle'=60,scale=0.6]}-] (4.2,1.2)--(4.5,1.3) node [right] {$\Sigma$};
   \draw (1,1.7) node [above] {$\mathbb{R}^{1,d-1}$};
   \draw (3.5,0.7) node [above] {$A$};
  \end{tikzpicture}
\end{minipage}
\hspace*{2cm}
\begin{minipage}{0.3\hsize} 
  \centering
  \begin{tikzpicture}[scale=1.2]
   \draw (0.2,0.4)--(2.0,0.4)--(3.6,1.6)--(1.8,1.6)--cycle;
   \draw[-{Stealth[angle'=60,scale=0.8]}] (1.8, 1.6)--(1.8, 2.5) node [right] {$t$};
   \draw(0.3,2.2) node [right] {$\mathcal{D}^{(d-1)}$};
   \filldraw[fill=orange,opacity=0.3,draw=none] (0.2,-0.4)--(1.8,0.8)--(1.8,2.4)--(0.2,1.2)--cycle;
   \draw (1.266,0.4)--(0.2,-0.4)--(0.2,1.2)--(1.8,2.4);
   \draw (1.8,1.6)--(1.8,2.4);
   \draw[dashed,opacity=0.5] (1.266,0.4)--(1.8,0.8)--(1.8,1.6);
   \draw[fill=cyan!40, thick] (0.45,0.585) arc [start angle = -118, end angle = 62, x radius=1.25cm, y radius=0.5cm];
   \draw (1.5,0.7) node [above] {$A$};
  \end{tikzpicture}
\end{minipage}
\caption{[Left] A dimension-$p$ conformal defect $\mathcal{D}^{(p)}$  for $p< d-1$ in Lorentzian flat spacetime. The spherical subsystem $A$ of radius $R$ surrounds the defect. [Right] A codimension-one defects $\mathcal{D}^{(d-1)}$ as a boundary. The subsystem $A$ intersects with the defect.}
\label{fig:setup}
\end{figure}

To define an entanglement entropy, we choose a time slice at $t=0$ in the Minkowski spacetime $\BR^{1,d-1}$ with the metric
\begin{align}
    \begin{aligned}
        \d s^2
            &=
                \eta_{\mu\nu} \, \d x^{\mu} \, \d x^{\nu} \\
            &=
                -\d t^2 + \delta_{ij} \, \d x^i\,\d x^j \qquad (i,j=1,\cdots,d-1) \ ,
    \end{aligned}   
\end{align}
and divide the time slice into two complementary regions $A$ and $\bar A$ separated by the codimension-two surface $\Sigma$ called the entangling surface (see figure \ref{fig:setup}).
The entanglement entropy of the region $A$ is defined as the von Neumann entropy of the reduced density matrix $\rho_A:= \tr_{\bar A}\left[\,\rho\,\right]$,
\begin{align}
    S_A 
        := 
        - \tr_A\left[\,\rho_A\,\log\rho_A\,\right] \ .
\end{align}
While this definition is valid in any quantum system, it is more convenient to employ the following representation as an alternative definition of entanglement entropy in quantum field theories \cite{Callan:1994py,Holzhey:1994we,Calabrese:2004eu}
 (see also \cite{Calabrese:2009qy,Nishioka:2009un,Casini:2009sr,Solodukhin:2011gn,Rangamani:2016dms,Nishioka:2018khk} for reviews):
\begin{align}\label{EE_def_QFT}
    S_A
        :=
            \lim_{n\to 1} \frac{1}{1-n}\,\log\,\frac{Z\left[ \CM_n\right]}{\left( Z\left[ \CM_1\right]\right)^n} \ ,
\end{align}
where $Z\left[ \CM_n\right]$ is the Euclidean partition function on the $n$-fold cover $\CM_n$ of $\BR^d$ with the conical singularity around $\Sigma$.

Let $\Sigma$ be a $(d-2)$-dimensional sphere of radius $R$ at $t=0$:
\begin{align}\label{entangling_surface}
    \Sigma
        =
            \left\{\,  x^\mu \in \BR^{1,d-1}~ \Big|~
            x^0=t=0,~ (x^1)^2+\cdots+(x^{d-1})^2=R^2
            \,\right\} \ .
\end{align}
Having the defect entropy in mind, we located a planar conformal defect $\CD^{(p)}$ at 
\begin{align}\label{planar_defect_config}
    \CD^{(p)}
        =
            \left\{\,  x^\mu \in \BR^{1,d-1}~ \Big|~
            x^{p}=\cdots=x^{d-1}=0
            \,\right\} \ .
\end{align}
For $p=d-1$, we restrict our attention to the boundary case (see figure 1 of \cite{Kobayashi:2018lil} for the interface setup).
Denoting the entanglement entropies in the DCFT and the ambient CFT by $S^{(\text{DCFT})}$ and $S^{(\text{CFT})}$, respectively, we define the defect entropy for $p \le d-2$ as
\begin{align}
    S_{\text{defect}}
        :=
            S^{(\text{DCFT})} - S^{(\text{CFT})} \ ,
\end{align}
and the boundary entropy for $p=d-1$ as
\begin{align}
    S_{\text{bdy}}
        :=
            S^{(\text{BCFT})} - \frac{1}{2}\,S^{(\text{CFT})} \ .
\end{align}
The defect and boundary entropies measure the additional contribution to the entanglement entropy due to the existence of a defect.
The defect entropy exhibits the UV divergences and should be regularized and renormalized appropriately.
By introducing the cutoff $\epsilon$, the defect entropy can be shown to have a structure of the UV divergence similar to that of the entanglement entropy in a $p$-dimensional CFT (see e.g., \cite{Jensen:2013lxa,Kobayashi:2018lil}):
\begin{align}\label{defect_entropy_divergence}
    S_{\text{defect}}
        =
            \frac{c'_{p-2}}{\epsilon^{p-2}} + \frac{c'_{p-4}}{\epsilon^{p-4}} +
            \cdots +
            \begin{cases}
                (-1)^\frac{p}{2}\,B'\,\log\,\epsilon + \cdots \quad &(p:\text{even}) \ ,\\
                 (-1)^\frac{p-1}{2}\,D' \quad &(p:\text{odd}) \ .
            \end{cases}          
\end{align}
While the coefficients $c_i' \, (i=p-2,\,p-4,\,\cdots)$ are scheme-dependent under the rescaling of the cutoff $\epsilon$, the coefficients $B', D'$ are universal in the sense that they are scheme-independent in DCFTs.
We will compute the defect entropy using the Ryu-Takayanagi formula in our holographic model in section \ref{subsec:Hol_entropy}.

Next, we turn to the defect free energy which is defined through the sphere free energy $\log\,Z[\,\BS^{d}\,]$ with the Euclidean partition function $Z[\,\BS^{d}\,]$ on $\BS^{d}$.
It follows from \eqref{EE_def_QFT} that in CFTs without defects the sphere free energy is related to the entanglement entropy for a spherical entangling surface by the relation:
\begin{align}\label{entropy_free_energy}
    S^{(\text{CFT})}
        =
            \log\,Z_{\text{CFT}}[\,\BS^{d}\,] \ ,
\end{align}
where the equality holds up to UV divergences \cite{Casini:2011kv}.

\begin{figure}[t]
   \centering
   \begin{tikzpicture}[scale=1.8, >=stealth]
    \draw (0,0) circle [radius = 1cm];
    \draw[dashed, orange] (1,0) arc [start angle=0, end angle=180, x radius = 1cm, y radius= 0.25cm];
    \draw[thick, orange] (1,0) arc [start angle=0, end angle=-180, x radius = 1cm, y radius= 0.25cm];
    \draw[-{Stealth[angle'=60,scale=0.8]}] (-0.25,0.075) arc [start angle=120, end angle=300, x radius = 5mm, y radius= 1mm];
    \draw (-0.8,0.8) node [left] {$\mathbb{S}^{d}$};
    \draw (-0.2,-0.3) node [below] {$\mathcal{D}^{(p)}$};
    \draw[-{Stealth[angle'=60,scale=0.8]}] (-0.5,-0.5) arc [start angle=-90, end angle=-180, x radius=2.0mm, y radius=3.0mm];
    \draw (0.15,-0.05) node [right=0.15cm] {$\BS^{p}$};
    \end{tikzpicture}
    \caption{A conformal defect $\mathcal{D}^{(p)}$ on $\mathbb{S}^{d}$, which wraps $\BS^p$ on the equator.
    }
    \label{fig:sphere}
\end{figure}
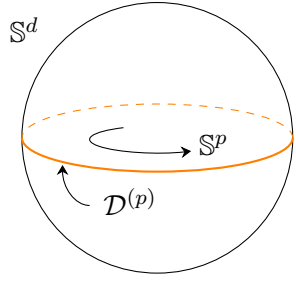

To define a DCFT on $\BS^{d}$, we use the conformal map \eqref{polar_coordinate} from $\BR^d$ to $\BH^{p+1} \times \BS^{d-p-1}$ followed by another conformal map from $\BH^{p+1} \times \BS^{d-p-1}$ to $\BS^d$.
For the latter, we employ the global coordinates for $\BH^{p+1}$ in the metric \eqref{hyperbolic_coordinate} as
\begin{align}
    \d s^2_{\BH^{p+1} \times \BS^{d-p-1}}
        =
            \d w^2 + \sinh^2w\,\d\Omega^2_{p} + \d\Omega^2_{d-p-1} \ ,
\end{align}
where the defect $\CD^{(p)}$ wraps $\BS^{p}$ at $w=\infty$.
By the coordinate transformation $\sinh w = \cot\varphi$, the metric becomes
\begin{align}
    \d s^2_{\BH^{p+1} \times \BS^{d-p-1}}
        =
            \frac{1}{\sin^2\varphi}\,
            \left(\,
            \d \varphi^2 + \cos^2\varphi\,\d\Omega^2_{p} + \sin^2\varphi\,\d\Omega^2_{d-p-1}
            \,\right) \ ,
\end{align}
which is conformally equivalent to $\BS^{d}$ with the metric
\begin{align}
    \d s^2_{\BS^{d}}
        =
            \d \varphi^2 + \cos^2\varphi\,\d\Omega^2_{p} + \sin^2\varphi\,\d\Omega^2_{d-p-1} \ .
\end{align}
After the sequence of the conformal maps, the planar defect $\CD^{(p)}$ on $\BR^d$ is mapped to $\BS^{p}$ at $\varphi=0$ on $\BS^d$ (see figure \ref{fig:sphere}).\footnote{
One can verify that this is the metric of a $d$-sphere $\BS^{d}$ by parameterizing $\BS^{d}$ in flat space $\BR^{d+1}$ as follows:
\begin{align}
    \left\{\
    \begin{aligned}
        x_0 &= \cos\varphi\,\cos\varphi_0 \ , \\
        x_1 &= \cos\varphi\,\sin\varphi_0\,\cos\varphi_1 \ , \\
        x_2 &= \cos\varphi\,\sin\varphi_0\,\sin\varphi_1\,\cos\varphi_2 \ , \\
        &\vdots \\
        x_{p-1} &= \cos\varphi\,\sin\varphi_0\,\cdots\,\sin\varphi_{p-2}\,\cos\varphi_{p-1} \ , \\
        x_{p} &= \cos\varphi\,\sin\varphi_0\,\cdots\,\sin\varphi_{p-2}\,\sin\varphi_{p-1} \ ,
    \end{aligned}
    \right.
    &&
    \left\{\
    \begin{aligned}
        x_{p+1} &= \sin\varphi\,\cos\varphi_{p+1} \ , \\
        x_{p+2} &= \sin\varphi\,\sin\varphi_{p+1}\,\cos\varphi_{p+2} \ , \\
        x_{p+3} &= \sin\varphi\,\sin\varphi_{p+1}\,\sin\varphi_{p+2}\,\cos\varphi_{p+3} \ , \\
        &\vdots \\
        x_{d-1} &= \sin\varphi\,\sin\varphi_{p+1}\,\cdots\,\sin\varphi_{d-2}\,\cos\varphi_{d-1} \ , \\
        x_{d} &= \sin\varphi\,\sin\varphi_{p+1}\,\cdots\,\sin\varphi_{d-2}\,\sin\varphi_{d-1} \ .
    \end{aligned}
    \right.
\end{align}
}

Similarly to the defect entropy, the additional contributions to the sphere free energy due to the defect can be quantified by the defect free energy 
\begin{align}\label{defect_free_energy}
    \log\,\langle\,\CD^{(p)}\,\rangle
        :=
            \log\,Z_{\text{DCFT}}[\,\BS^{d}\,] - \log\,Z_{\text{CFT}}[\,\BS^{d}\,] \ ,
\end{align}
for $p < d-1$, and
\begin{align}
    \log\,\langle\,\CD^{(d-1)}\,\rangle|_{\text{BCFT}}
        :=
            \log\,Z_{\text{BCFT}}[\,\BH\BS^{d}\,] - \frac{1}{2}\,\log\,Z_{\text{CFT}}[\,\BS^{d}\,] \ ,
\end{align}
for BCFTs, where $\BH\BS^{d}$ is the $d$-dimensional hemisphere.
The defect free energy also has the UV divergences
\begin{align}\label{defect_energy_divergence}
    \log\,\langle\,\CD^{(p)}\,\rangle
        =
            \frac{c_{p}}{\epsilon^{p}} + \frac{c_{p-2}}{\epsilon^{p-2}} +
            \cdots +
            \begin{cases}
                (-1)^\frac{p}{2}\,B\,\log\,\epsilon + \cdots \quad &(p:\text{even}) \ ,\\
                 (-1)^\frac{p-1}{2}\,D \quad &(p:\text{odd}) \ ,
            \end{cases}          
\end{align}
where the coefficients $c_i \, (i=p,\,p-2,\,\cdots)$ are scheme-dependent while $B, D$ are universal constants different from $B', D'$ in general.
While the defect entropy differs from the defect free energy even up to UV divergences, in contrast to the case of the absence of the defect in \eqref{entropy_free_energy}, they are related by the following relation which holds up to UV divergences \cite{Kobayashi:2018lil} (see also \cite{Lewkowycz:2013laa} for $p=1$):\footnote{This relation is derived by using the dimensional regularization and assuming that there are no conformal anomalies.}
\begin{align}\label{universal}
    S_{\text{defect}}
        =
            \log\,\langle\,\CD^{(p)}\,\rangle
            -
            \frac{2\,(d-p-1)\,\pi^{\frac{d}{2}+1}}{\sin\left(\frac{\pi\, p}{2}\right)\,d\,\Gamma\left(\frac{p}{2}+1\right)\,\Gamma\left(\frac{d-p}{2}\right)}\,a_T \ ,
\end{align}
where $a_T$ is the coefficient of the one-point function of the stress tensor \eqref{one_point_stress}.

We conclude this subsection with a few remarks:
\begin{itemize}
    \item The second term in the right hand side of the relation \eqref{universal} vanishes for $p=d-1$, thus the boundary entropy coincides with the defect free energy (up to UV divergences):
        \begin{align}\label{boundary_entropy}
            S_{\text{bdy}}
                =
                     \log\,Z_{\text{BCFT}}[\,\BH\BS^{d}\,] - \frac{1}{2}\,\log\,Z_{\text{CFT}}[\,\BS^{d}\,] \ .
        \end{align}
        This relation simplifies the computation of the boundary entropy, which is generally more difficult to evaluate than the sphere free energy.
    \item In the absence of conformal anomalies both in the ambient CFT and on the defect, the sphere partition function equals to that on $\BH^{p+1} \times \BS^{d-p-1}$:
        \begin{align}\label{partition_function}
            Z_\text{DCFT}[\,\BS^d\,]
                =
                Z_\text{DCFT}[\,\BH^{p+1} \times \BS^{d-p-1}\,] \ .
        \end{align}
    We will use this relation in calculating the defect free energy holographically in section \ref{subsec:Hol_free_energy}.
\end{itemize}

\subsection{Defect $\CC$-theorem}\label{ss:defect_C}

Searching for constraints on RG flows is one of the central issues in quantum field theory.
Such a flow interpolates between a CFT at a UV fixed point and another CFT at an IR fixed point.
In the Wilsonian picture, renormalization is interpreted as a sequence of coarse graining, implying the irreversibility of RG flows.
The $\CC$-theorem substantiates this perspective by asserting the existence of a monotonic function, a so-called $\CC$-function which measures the effective number of degrees of freedom in a QFT.

This paradigm originated from the Zamolodchikov's $c$-theorem in two dimensions \cite{Zamolodchikov:1986gt} and has since been extended to higher dimensions.
In even dimensions, the type-$A$ conformal anomaly is conjectured to be a $C$-function \cite{Cardy:1988cwa,Myers:2010xs,Myers:2010tj}.
In particular, the four-dimensional $C$-theorem, known as the $a$-theorem, was proven in \cite{Komargodski:2011vj}.
In odd dimensions, where no conformal anomalies exist, it has been conjectured that the universal part of the sphere free energy $F := (-1)^{\frac{d-1}{2}}\log\,Z[\,\BS^{d}\,]$ plays a role of a $C$-function \cite{Jafferis:2011zi,Klebanov:2011gs}.
This statement, known as the $F$-theorem, was proved in three dimensions by using the key relation \eqref{entropy_free_energy} between the sphere free energy and the entanglement entropy \cite{Casini:2012ei}.
See also \cite{Casini:2004bw,Komargodski:2011xv,Casini:2023kyj,Hartman:2023ccw,Hartman:2023qdn} for alternative proofs of the $\CC$-theorems in $d\le 4$ dimensions.

While the $\CC$-theorems take quite different forms in even and odd dimensions, the type-$A$ anomaly and the sphere free energy $F$ can be interpolated by the following function:
\begin{align}\label{C_function}
    \tilde{F}
        :=
            \sin\left(\frac{\pi\,d}{2}\right)\,\log\,Z[\,\BS^{d}\,] \ .
\end{align}
The generalized $F$-theorem recasts the $\CC$-theorems in diverse dimensions as a statement that $\tilde F$ serves as a $\CC$-function, i.e., $\tilde F$ is positive and decreases along any RG flow \cite{Giombi:2014xxa}:
\begin{align}\label{F_theorem}
    \tilde{F}_{\text{UV}} 
        \ge 
            \tilde{F}_{\text{IR}} \ .
\end{align}
Although the generalized $F$-theorem in $d>4$ has not been proven yet, there is accumulating evidence supporting its validity \cite{Myers:2010tj,Jafferis:2012iv,Fei:2014yja,Cordova:2015fha,Heckman:2015axa}.

A tempting generalization of the $\CC$-theorem is to QFTs with defects, seeking a $\CC$-function that decreases under either ambient RG flows or defect-localized RG flows.
In what follows, we focus on the latter triggered by defect-localized operators, while keeping the ambient CFT fixed.\footnote{See  \cite{Green:2007wr,Herzog:2019rke,Bianchi:2019umv,Sato:2020upl,Herzog:2021hri,Shachar:2024ubf} for the works on $\CC$-theorems under ambient RG flows in the presence of defects.}
In QFTs with a $p$-dimensional defect $\CD^{(p)}$, the defect free energy $\log\,\langle\,\CD^{(p)}\,\rangle$ plays a similar role as the sphere free energy.
Motivated by this analogy with \eqref{C_function} in mind, it is natural to define
\begin{align}\label{defect_C_function}
    \tilde{D}
        :=
            \sin\left(\frac{\pi\,p}{2}\right)\,\log\,\langle\,\CD^{(p)}\,\rangle \ ,
\end{align}
and to conjecture that $\tilde \CD$ be a $\CC$-function that decreases along any defect-localized RG flow:
\begin{align}\label{defect_C_theorem}
    \tilde{D}_{\text{UV}} 
        \ge 
            \tilde{D}_{\text{IR}} \ .
\end{align}
This statement was proposed in \cite{Kobayashi:2018lil} and is termed the defect $\CC$-theorem.
Note that $\tilde{D}$ coincides with $B$ and $D$ in \eqref{defect_energy_divergence} up to a positive factor
\begin{align}
    \tilde{D}
        =
        \begin{cases}
            \frac{\pi}{2}\,B \qquad &(p:\text{even}) \ ,
            \\
            D \qquad &(p:\text{odd}) \ ,
        \end{cases}
\end{align}
where $B$ is the type-$A$ conformal anomaly on a defect.
While $\tilde{F}$ is positive in known examples of unitary CFTs, $\tilde{D}$ is not necessarily positive even in unitary DCFTs \cite{Nozaki:2012qd,Jensen:2015swa,Fursaev:2016inw}.
Thus far, the conjecture of the defect $\CC$-theorem has successfully passed several nontrivial checks and has unified previously known theorems.\footnote{Defect entropies do not necessarily decrease under defect-localized RG flows unless for $p=d-1$ \cite{Kumar:2017vjv,Kobayashi:2018lil,Rodgers:2018mvq}.} 
Non-perturbative proofs in a field theoretic framework have been established for $p=1$ \cite{Cuomo:2021rkm}, $p=2$ (the so-called $b$-theorem) \cite{Jensen:2015swa,Shachar:2022fqk} , and $p=4$ \cite{Wang:2021mdq}.
In the case of BCFTs where $p=d-1$, $\tilde D$ coincides with the boundary entropy through the relation \eqref{boundary_entropy}, thus the defect $\CC$-theorem reduces to the $g$-theorem for $d=2$ \cite{Friedan:2003yc,Cuomo:2021rkm,Casini:2016fgb,Casini:2022bsu,Harper:2024aku} (see also \cite{Casini:2018nym} for the $g$-theorem for $d=3$).
There are also extensive studies for the proposal from both holographic \cite{Affleck:1991tk,Yamaguchi:2002pa,Takayanagi:2011zk,Fujita:2011fp,Estes:2014hka} and field theoretic perspectives \cite{Gaiotto:2014gha,Wang:2020xkc,Giombi:2020rmc,Nishioka:2021uef,Sato:2021eqo,Chalabi:2021jud,Yuan:2022oeo,Casini:2023kyj,Santilli:2023fuh,Ge:2024hei}.

In section \ref{subsec:Hol_RG}, we will give a holographic proof of the defect $\CC$-theorem in our model of holographic DCFTs, where the defect-localized RG flow is described by a localized scalar field on the end-of-the-world brane.

\section{Holographic models of defect CFTs}
\label{sec:Hol_DCFT}
We now turn to the holographic description of a DCFT with a $p$-dimensional conformal defect $\CD^{(p)}$.
Section \ref{ss:AdS_BCFT} reviews the AdS/BCFT model \cite{Takayanagi:2011zk,Fujita:2011fp}, in which the bulk AdS space has a boundary not only at the asymptotic infinity but also on an End-of-the-World (EoW) brane anchored at the boundary of the dual BCFT.
In section \ref{ss:AdS_DCFT}, we extend the AdS/BCFT model and propose a holographic model of DCFTs by introducing an EoW brane whose boundary lies on the infinitesimal tubular neighborhood of $\CD^{(p)}$.
Unlike the AdS/BCFT model, where the Neumann boundary condition is imposed at the EoW brane, our construction employs the Dirichlet boundary condition.
In this setup, we compute the defect entropy and defect free energy holographically and read off the coefficient $a_T$.
In section \ref{subsec:Hol_RG}, we show that the defect $\CC$-function is always non-negative in our model and prove the defect $\CC$-theorem for defect-localized RG flows that are described holographically by a scalar field localized on the brane.

\subsection{Review of AdS/BCFT model}\label{ss:AdS_BCFT}

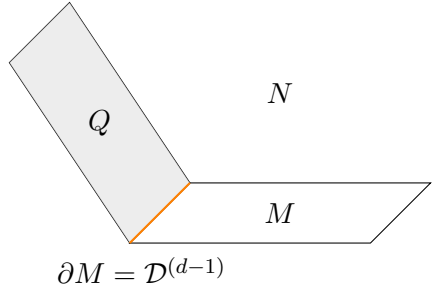
\begin{figure}[t]
    \centering
    \begin{tikzpicture}[scale=0.8]
        \draw (0,0) -- (4,0) -- (5,1) -- (1,1) -- cycle;
        \draw (0,0) -- (-2,3) -- (-1,4) -- (1,1) -- cycle;
        \fill[gray!20,opacity=0.7] (0,0) -- (-2,3) -- (-1,4) -- (1,1) -- cycle;
        \draw[orange,thick] (0,0) -- (1,1);
        \node at (2.5,2.5) {$N$};
        \node at (2.5,0.5) {$M$};
        \node at (-0.5,2) {$Q$};
        \node at (0.2,-0.45) {$\partial M =\CD^{(d-1)}$};
    \end{tikzpicture}
    \caption{In the AdS/BCFT model, the bulk spacetime $N$ ends at the AdS boundary $M$ and the EoW brane $Q$ with $\partial Q = \partial M$. The boundary $\partial M$ of the BCFT is regarded as a codimension-one defect $\CD^{(d-1)}$.}
    \label{fig:AdS/BCFT}
\end{figure}

In the AdS/BCFT model \cite{Takayanagi:2011zk,Fujita:2011fp}, the bulk $(d+1)$-dimensional AdS space $N$ is bounded at the asymptotic boundary by a $d$-dimensional space $M$ where the dual BCFT lives and the other codimension-one hypersurface $Q$ with boundary $\partial Q = \partial M$ (see figure \ref{fig:AdS/BCFT}).
To terminate the bulk space $N$ on $Q$, one can locate the EoW brane with a brane localized matter:\footnote{We drop the Gibbons–Hawking term for $M$, since it does not affect the following analysis.}
\begin{align}
    I
        =
            -\frac{1}{16\pi\, G_N}\int_N \d^{d+1}X\,\sqrt{g}\,\left( \CR - 2\Lambda\right) 
            -
            \frac{1}{8\pi\, G_N}\int_Q \d^d y\,\sqrt{h}\, \CK  + \frac{1}{8\pi\, G_N}\,I_Q\ ,
\end{align}
where $g_{AB}$ and $h_{ab}$ are the metrics on $N$ and $Q$, respectively, and $\CR$ and $\CK$ are the Ricci scalar and the trace of the extrinsic curvature.
$I_Q$ denotes the localized matter action on $Q$, where we choose the normalization so as to simplify the equations of motion.
Using the outward-pointing unit vector $n^A$ normal to $Q$, the induced metric $h$ is written as
\begin{align}
    h_{ab}
        =
            e^A_a\,e^B_b\,(g_{AB} - n_A\,n_B) \ ,
\end{align}
where $e^A_a = \frac{\partial X^A}{\partial y^a}$.
The extrinsic curvature $\CK_{ab}$ is defined by
\begin{align}
    \CK_{ab}
        =
            e_a^{~K}\,e_b^{~L}\,\nabla_K n_L \ ,
\end{align}
and the trace of the extrinsic curvature $\CK$ is written as
\begin{align}
    \CK 
        =
            h^{ab}\,\CK_{ab}
        =
            \nabla_A n^A \ .
\end{align}
When the hypersurface $Q$ is defined as a constant slice of a function $f$, the normal vector is given by
\begin{align}
    n_M
        =
            \frac{\partial_M f}{\sqrt{g^{MN}\,\partial_M f\,\partial_N f}} \ .
\end{align}
In this case, the variation of the action becomes \cite{brown1993quasilocal-0e8} (see also \cite{harlow2020covariant-144})
\begin{align}
    \delta I
        =
            -
            \frac{1}{8\pi\, G_N}\int_Q \d^d y\,\sqrt{h}\,\left( \CK\,h^{ab} - \CK^{ab} + T_Q^{ab}\right)\,\delta h_{ab} \ ,
\end{align}
where $T_Q$ is the stress tensor for the localized matter:
\begin{align}
    T_Q^{ab}
        =
            -\frac{2}{\sqrt{h}}\,\frac{\delta I_Q}{\delta h_{ab}} \ .
\end{align}
While the Dirichlet boundary condition is set on the asymptotic AdS boundary as in the standard AdS/CFT setup, the AdS/BCFT model imposes the Neumann boundary condition on the EoW brane at $Q$,\footnote{See \cite{Miao:2018qkc} for the holographic BCFT with the Dirichlet boundary condition and \cite{Miao:2017gyt,Chu:2017aab} for another boundary condition.} leading to the equation of motion:
\begin{align}\label{Neumann_bc}
    \CK_{ab} - \CK\,h_{ab}
        =
            T_{Q\,ab} \ .
\end{align}
For a brane with tension, $I_Q = T\,\int_Q \d^d y\,\sqrt{h}$,
the Neumann boundary condition \eqref{Neumann_bc} reduces to
\begin{align}
    \CK_{ab}
        =
            (\CK - T)\,h_{ab} \ .
\end{align}
By taking the trace of both sides, we find
\begin{align}
    \CK
        =
            \frac{d}{d-1}\,T \ ,
\end{align}
which in turn implies that the extrinsic curvature is proportional to the induced metric on $Q$:
\begin{align}\label{EOM_K_h}
    \CK_{ab}
        =
            \frac{T}{d-1} \,h_{ab} \ .
\end{align}
In the AdS/BCFT model, the configuration of the EoW brane is dynamically determined so as to satisfy the equation \eqref{EOM_K_h}.

\subsection{AdS/DCFT models with Dirichlet boundary conditions}\label{ss:AdS_DCFT}
We move onto identifying the holographic description of DCFTs.
We start with a $p$-dimensional planar defect $\CD^{(p)}$ on $\BR^d$ defined in \eqref{planar_defect}. 
As described in section \ref{ss:coordinates}, it is convenient to use the polar coordinates \eqref{polar_coordinate} of $\BR^d$ and locate the defect at the origin $r=0$.
To describe the dual geometry, we use the Poincar\'e metric of the AdS$_{\d+1}$ space:
\begin{align}\label{Poincare_metric}
    \d s^2
        =
            \frac{L^2}{z^2} \left[ \, \d z^2 + \d \hat{x}^2_{\hat a} + \d r^2  + r^2\,\d \Omega_{d-p-1}^2 \, \right] \ .
\end{align}
In the DCFT side, we performed the conformal map from $\BR^d$ to $\BH^{p+1}\times \BS^{d-p-1}$ with the metric \eqref{hyperbolic_coordinate} which manifests the $\SO(1,p+1) \times \SO(d-p)$ symmetry.
In the bulk side, such a conformal map is realized by the coordinate transformation \cite{Jensen:2013lxa}
\begin{align}\label{coord_transf}
    z
        =
            \frac{Z}{\cosh \rho} \ , \qquad
    r
        =
            Z\,\tanh\rho \ .
\end{align}
This results in the hyperbolic slicing of the AdS$_{d+1}$ space: 
\begin{align}\label{metric_AdS_AdSslice}
    \d s^2
        =
            L^2\left[ 
                \d \rho^2 
                +
                \cosh^2\rho\,\frac{\d Z^2 + \d \hat x^2_{\hat a}}{Z^2}
                +
                \sinh^2\rho\,\d\Omega_{d-p-1}^2
                \right] \ ,
\end{align}
where $\rho\in [0,\infty),~Z\in [0,\infty)$ for $p<d-1$ and $\rho\in \BR$ for $p=d-1$.

\begin{figure}[t]
    \centering
    \begin{tikzpicture}[scale=0.8]

    \draw (0,0) -- (7.0,0) -- (9.0,1.5) -- (2.0,1.5) -- cycle;
    \draw (4.0,0.75) arc (180:360:0.5 and 0.2);
    \draw[dashed] (4.0,0.75) arc (180:0:0.5 and 0.2);
    \draw (7.1,3.3) -- (5.0,0.75);
    \draw (1.9,3.3) -- (4.0,0.75);
    \draw (1.9,3.3) arc (180:360:2.6 and 0.4);
    \draw[dashed] (1.9,3.3) arc (180:0:2.6 and 0.4);
    \draw (4.5,0.75) -- (4.85,0.5);
    \draw[-{Stealth[angle'=60,scale=0.8]}] (4.5,0.75) -- (4.5,4.0);
    \draw[-{Stealth[angle'=60,scale=0.8]}] (4.5,0.75) -- (6.0,0.8);
    \filldraw[orange] (4.5,0.75) circle [x radius = 0.05, y radius = 0.05];
    \fill[gray!30, opacity = 0.4] (4.0,0.75) arc (180:360:0.5 and 0.2) -- (7.1,3.3) arc (360:180:2.6 and 0.4) -- (4.0,0.75);
    \node at (4.5,4.3) {$z$};
    \node at (6.3,0.8) {$r$};
    \node at (7.5,2.5) {$N$};
    \node at (7.8,1.1) {$M$};
    \node at (7.4,3.5) {$Q$};
    \node at (3.7,0.5) {$N_{\epsilon}$};
    \node at (5.1,0.3) {$\CD^{(p)}$};

    \begin{scope}[shift={(9.0,0)}]
        \draw (0,0) -- (7.0,0) -- (9.0,1.5) -- (2.0,1.5) -- cycle;
        \draw (6.9,3.3) -- (4.5,0.75);
        \draw (2.1,3.3) -- (4.5,0.75);
        \draw (2.1,3.3) arc (180:360:2.4 and 0.4);
        \draw[dashed] (2.1,3.3) arc (180:0:2.4 and 0.4);
        \draw (4.5,0.75) -- (4.85,0.5);
        \draw[-{Stealth[angle'=60,scale=0.8]}] (4.5,0.75) -- (4.5,4.0);
        \draw[-{Stealth[angle'=60,scale=0.8]}] (4.5,0.75) -- (6.0,0.8);
        \filldraw[orange] (4.5,0.75) circle [x radius = 0.05, y radius = 0.05];
        \fill[gray!30, opacity = 0.4] (4.5,0.75) -- (6.9,3.3) arc (360:180:2.4 and 0.4) -- (4.5,0.75);
        \node at (4.5,4.3) {$z$};
        \node at (6.3,0.8) {$r$};
        \node at (7.5,2.5) {$N$};
        \node at (7.8,1.1) {$M$};
        \node at (7.2,3.5) {$Q$};
        \node at (5.1,0.3) {$\CD^{(p)}$};
    \end{scope}
    
    \end{tikzpicture}
    \caption{[Left] Holographic description of DCFTs with a defect $\CD^{(p)}$. The hypersurface $Q$ ends on $N_{\epsilon}$ at the AdS boundary $M$. The bulk geometry $N$ is bounded by $Q$ and $M$. [Right] Taking the $\epsilon \to 0$ limit, $N_{\epsilon}$ shrinks to $\CD^{(p)}$.}
    \label{fig:holography}
\end{figure}
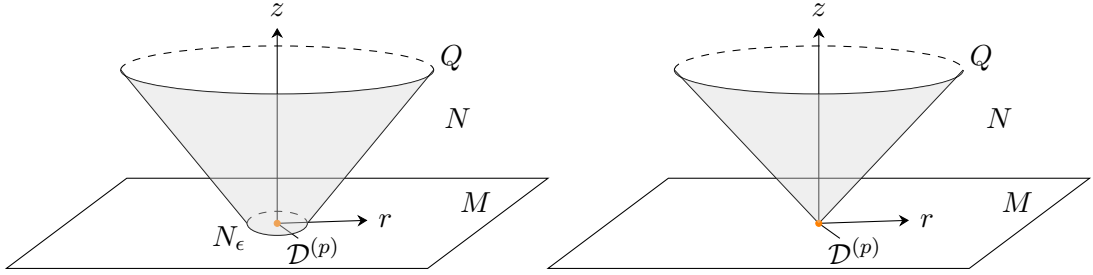

We now make use of the description of a defect as a boundary condition for ambient fields on the tubular neighborhood $N_\epsilon$ introduced in section \ref{ss:coordinates}.
In this picture, a DCFT can be viewed as a BCFT on $M$ with boundary $\partial M = N_\epsilon$, and the AdS/BCFT model may be employed for the gravitational dual.
Namely, we introduce the codimension-one hypersurface $Q$ in the AdS$_{d+1}$ space with the boundary $\partial Q = \partial M = N_{\epsilon}$ and take the $\epsilon \to 0$ limit for the infinitesimal tubular neighborhood $N_{\epsilon}$ (see figure \ref{fig:holography}).
Before going into the detail, we note that a similar setup arises in the cone holography \cite{Miao:2021ual,Cui:2023gtf}, where the bulk geometry inside the cone-like hypersurface $Q$ describes the defect theory but not the ambient CFT. 
In contrast, the bulk is taken to be the exterior of $Q$ to capture the full DCFT in our model.

We illustrate our approach by the EoW brane model with constant tension, whose action is given by
\begin{align}\label{HDCFT_action}
    I
        &=
            -\frac{1}{16\pi\, G_N}\int_N \d^{d+1}X\,\sqrt{g}\,\left( \CR  + \frac{d(d-1)}{L^2}\right) 
            -
            \frac{1}{8\pi\, G_N}\int_Q \d^d y\,\sqrt{h}\,\left( \CK - T\right) \ .
\end{align}
To realize the $\SO(1,p+1) \times \SO(d-p)$ symmetry of the dual DCFT with $\CD^{(p)}$, we assume that the EoW brane lies at a constant $\rho$ slice, i.e, the hypersurface $Q$ is located at $\rho=\rho_{\ast}$ for constant $\rho_\ast$. (We take the $\epsilon \to 0$ limit for $N_\epsilon$ from the beginning (see the right panel of figure \ref{fig:holography}).)
The bulk AdS$_{d+1}$ space is bounded within the region $\rho \in [\rho_\ast,\infty)$.
The induced metric on $Q$ becomes
\begin{align}\label{induced_metric}
    \d s^2_Q
        =
            h_{ab}\,\d y^a\,\d y^b
        =
            L^2\left[ 
                \cosh^2\rho_\ast\,\frac{\d Z^2 + \d \hat{x}^2_{\hat{a}}}{Z^2}
                +
                \sinh^2\rho_\ast\,\d\Omega_{d-p-1}^2
                \right] \ ,
\end{align}
from which the extrinsic curvature on $Q$ is calculated as
\begin{align}\label{extrinsic_curvature_brane}
    \begin{aligned}
        \CK_{ZZ}
            &=
                - \frac{\tanh\rho_\ast}{L}\,h_{ZZ} \ , \\
        \CK_{\hat a \hat a}
            &=
                - \frac{\tanh\rho_\ast}{L}\,h_{\hat a \hat a} \ , \\
        \CK_{ii}
            &=
                - \frac{\coth\rho_\ast}{L}\,h_{ii} \ .
    \end{aligned} 
\end{align}
where the $ii$ components represent the $\BS^{d-p-1}$ direction.
The trace of the extrinsic curvature is 
\begin{align}\label{extrinsic_curvature_trace}
    \CK
        =
            -\frac{(p+1)\tanh\rho_\ast + (d-p-1)\coth\rho_\ast}{L} \ . 
\end{align}
While the brane configuration is fixed by symmetry, it remains necessary to check if the ansatz is consistent with the boundary condition imposed on $Q$.

For the Neumann case, the extrinsic curvature on $Q$ has to satisfy the equations of motion \eqref{EOM_K_h}.
For the $ZZ$ and $\hat{a}\hat{a}$ components, comparing \eqref{extrinsic_curvature_brane} with \eqref{EOM_K_h} gives
\begin{align}\label{tension_rho_relation_ZZ_aa}
    T 
        =
            - (d-1)\,\frac{\tanh\rho_\ast }{L} \ .
\end{align}
For $p=d-1$, \eqref{EOM_K_h} is solved by \eqref{tension_rho_relation_ZZ_aa}, since there are no spherical directions. 
In this case, $\rho_\ast$ can takes any real value, and physical solutions with $T>0$ are obtained by choosing $\rho_\ast <0$, as in \cite{Takayanagi:2011zk,Fujita:2011fp}.
On the other hand, for $p< d-1$, the $ii$ components of \eqref{EOM_K_h} yields the additional condition
\begin{align}\label{tension_rho_relation_ii}
    T 
        =
            - (d-1)\,\frac{\coth\rho_\ast }{L} \ .
\end{align}
The two relations \eqref{tension_rho_relation_ZZ_aa} and \eqref{tension_rho_relation_ii} are compatible only when $\rho_{\ast}=\infty$, but this solution is unphysical because the bulk geometry collapses completely.
Hence, the Neumann boundary condition does not give physically reasonable solutions unless $p=d-1$.
This implies that the AdS/BCFT model with Neumann boundary condition cannot be extended to describe DCFTs of codimension greater than one.
 
To circumvent the above problem for $p< d-1$, we propose to employ the Dirichlet boundary condition which fixes the induced metric on $Q$ to be \eqref{induced_metric}.\footnote{
In the context of the cone holography, the other boundary conditions have been used; (i) imposing a mixed boundary condition, with the Neumann boundary condition along the sub-AdS ($\BH^{p+1}$) direction and the Dirichlet boundary condition along the spherical ($\BS^{d-p-1}$) direction \cite{Miao:2021ual}, (ii) localizing the $p$-form gauge fields in addition to the brane tension and imposing the Neumann boundary condition \cite{Cui:2023gtf}.
These boundary conditions allow physical brane solutions with positive tension for the cone holography, but yield brane solutions with negative tension in our model as our bulk region is outside of $Q$ and the sign of the extrinsic curvature is opposite to theirs.
}
With this prescription, the extrinsic curvature on $Q$ is no longer subject to the equation of motion \eqref{Neumann_bc}, hence we treat the tension $T$ and the position $\rho_\ast$ of the EoW brane as free parameters.

In what follows, we will consider physical solutions for $p<d-1$ with the positive brane tension $T>0$, and examine the defect entropy and defect free energy holographically in our model.

\subsubsection{Holographic defect entropy}\label{subsec:Hol_entropy}

Using our model, we compute the defect entropy defined in section \ref{subsec:entropy_free_energy}.
We begin with the Poincar\'e patch of the Lorentzian AdS$_{d+1}$ spacetime, obtained from \eqref{Poincare_metric} by the analytic continuation $\hat x^0 = \i\,t$.
On the AdS boundary $(z=0)$, we place a spherical entangling surface $\Sigma$ of radius $R$ and a planar defect $\CD^{(p)}$ as in \eqref{entangling_surface} and \eqref{planar_defect_config}, respectively.

\begin{figure}
    \centering
    \begin{tikzpicture}[scale=1.0]
        \draw (0,-0.5) -- (7.0,-0.5) -- (9.0,2.0) -- (2.0,2.0) -- cycle;
        \fill [fill=cyan!50, opacity=0.7](4.5,0.75) circle [x radius = 1.0, y radius = 0.4];
        \draw [draw=cyan!50, opacity=0.7](3.5,0.75) arc (180:360:1.0 and 0.4);
        \draw [draw=cyan!50, opacity=0.7, dashed] (3.5,0.75) arc (180:0:1.0 and 0.4);
        \draw (6.9,3.3) -- (4.5,0.75);
        \draw (2.1,3.3) -- (4.5,0.75);
        \draw (2.1,3.3) arc (180:360:2.4 and 0.4);
        \draw[dashed] (2.1,3.3) arc (180:0:2.4 and 0.4);
        \fill[gray!30, opacity = 0.4] (4.5,0.75) -- (6.9,3.3) arc (360:180:2.4 and 0.4) -- (4.5,0.75);

        \draw [black, thick](4.5,0.75) circle [x radius = 1.0, y radius = 0.4];
        
        \draw[fill=yellow!50, draw=yellow, opacity=0.5] (3.82,1.49) arc (133:180:1.0 and 1.0) arc (180:360:1.0 and 0.4) arc (0:47:1.0 and 1.0) arc (0:-180:0.68 and 0.15) -- cycle;
        
        \draw[draw=red, opacity=0.5] (3.5,0.75) arc (180:133:1.0 and 1.0);
        \draw[draw=red, opacity=0.5] (5.5,0.75) arc (0:47:1.0 and 1.0);
        
        \draw (3.82,1.49) arc (180:360:0.68 and 0.15) ;
        \draw[dashed] (3.82,1.49) arc (180:0:0.68 and 0.15);

        \draw (4.5,0.75) -- (5.2,0.25);
        \draw (5.45,1.1) -- (5.7,1.2);
        \draw (3.3,0.4) -- (3.6,0.58);
        \draw (8.0,5.0) -- (8.0,4.5) -- (9.5,4.5);
        \draw[-{Stealth[angle'=60,scale=0.8]}] (4.5,0.75) -- (4.5,4.0);
        \filldraw[orange] (4.5,0.75) circle [x radius = 0.05, y radius = 0.05];
        \node at (4.5,4.3) {$z$};
        \node at (7.5,2.5) {$N$};
        \node at (7.5,1.0) {$M$};
        \node at (7.2,3.5) {$Q$};
        \node at (5.5,0.1) {$\CD^{(p)}$};
        \node at (5.8,1.3) {$\gamma$};
        \node at (3.1,0.2) {$\Sigma$};
        \node at (8.75,4.75) {$t=0$};
    \end{tikzpicture}
    \caption{The RT surface $\gamma$ extends from the entangling surface $\Sigma$ to $Q$.}
    \label{fig:RT_surface}
\end{figure}
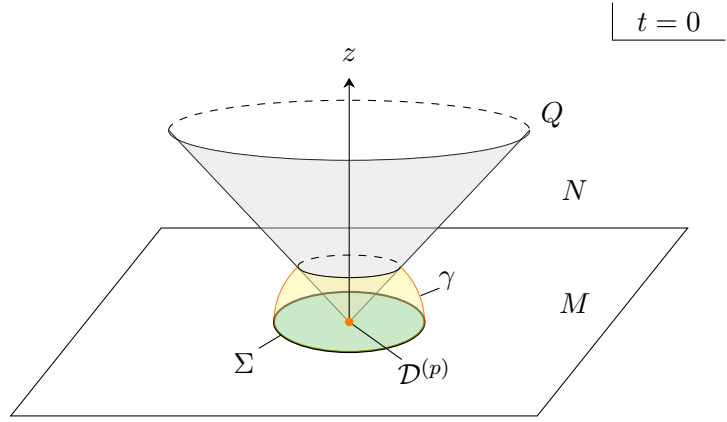

The entanglement entropy $S$ is calculated holographically by the Ryu-Takayanagi formula \cite{Ryu:2006bv,Ryu:2006ef}:
\begin{align}\label{RT_formula}
    S
        =
            \frac{\text{Area}(\gamma)}{4\,G_N} \ ,
\end{align}
where $\text{Area}(\gamma)$ is the area of the codimension-two minimal surface $\gamma$ called the Ryu-Takayanagi (RT) surface anchored on $\Sigma$.
To implement this prescription, it is convenient to switch from the Poincar\'e patch to the hyperbolic slicing of the AdS spacetime. 
By applying the coordinate transformation \eqref{coord_transf}, the Poincar\'e patch is mapped to the Lorentzian counterpart of the hyperbolic slicing \eqref{metric_AdS_AdSslice} \cite{Jensen:2013lxa}:
\begin{align}
    \d s^2
        =
            L^2\left[ 
                \d \rho^2 
                +
                \cosh^2\rho\,\frac{\d Z^2 -\d t^2 + \d r^2_{||} + r^2_{||}\d \Omega^2_{p-2}}{Z^2}
                +
                \sinh^2\rho\,\d\Omega_{d-p-1}^2
                \right] \ .
\end{align}
In these coordinates, the AdS boundary at $\rho=\infty$ becomes $\BH^{p+1} \times \BS^{d-p-1}$.
The defect $\CD^{(p)}$ is located at $Z=0$ while the entangling surface $\Sigma$ sits at
\begin{align}\label{RT_surface}
    t=0\ ,
    \qquad
    Z^2 + r^2_{||} = R^2 \ .
\end{align}
In the present case where the bulk geometry is locally AdS$_{d+1}$ spacetime, the RT surface $\gamma$ extends from $\Sigma$ into the bulk along the $\rho$ direction while keeping \eqref{RT_surface} and terminates on $Q$ at $\rho=\rho_\ast$ (see figure \ref{fig:RT_surface}).
Thus, the holographic entanglement entropy is calculated by\footnote{In the AdS/BCFT model, the edge of the RT surface on the EoW brane is dynamically determined by the Neumann boundary condition.
On the other hand, we fix the edge as the intersection between the RT surface in pure AdS and the EoW brane, and assume that the RT formula \eqref{RT_formula} remains valid for computing the entanglement entropy of the dual DCFT in our model.
We thank T.\,Takayanagi for valuable discussions on this point.
} 
\begin{align}
    \begin{aligned}
        S^{(\text{DCFT})}
            &=
                \frac{L^{d-1}}{4\,G_N}\,\text{Vol}(\BH^{p-1})\,\text{Vol}(\BS^{d-p-1})
                \int^{\infty}_{\rho_\ast}\,\d\rho\,\cosh^{p-1}\rho\,
                \sinh^{d-p-1}\rho \ ,
    \end{aligned}
\end{align}
where $\text{Vol}(\BH^{p-1})$ is the volume of the unit hyperbolic space and $\text{Vol}(\BS^{d-p-1})$ is the volume of the unit sphere given by
\begin{align}\label{regularized_volume}
    \text{Vol}(\BS^{d-p-1})
        =
            \frac{2\,\pi^\frac{d-p}{2}}{\Gamma\left(\frac{d-p}{2}\right)} \ .
\end{align}
By subtracting the CFT result $S^{(\text{CFT})} = S^{(\text{DCFT})}\big|_{\rho_\ast = 0}$, we obtain the defect entropy:
\begin{align}\label{Hol_defect_entropy}
    \begin{aligned}
        S_{\text{defect}}
            &=
                - \frac{L^{d-1}}{4\,G_N}\,\text{Vol}(\BH^{p-1})\,\text{Vol}(\BS^{d-p-1})
                \int_0^{\rho_\ast}\,\d\rho\,\cosh^{p-1}\rho\,
                \sinh^{d-p-1}\rho \\
            &=
                - \frac{L^{d-1}}{4\,G_N}\,\text{Vol}(\BH^{p-1})\,\text{Vol}(\BS^{d-p-1})\,
                \frac{1}{d-p}\,\tanh^{d-p}\rho_\ast\, 
                \cdot {}_2F_1 \left(\frac{d-p}{2}, \frac{d}{2}, \frac{d-p+2}{2}; \tanh^2\rho_\ast \right) \ .
    \end{aligned}   
\end{align} 

Let us check if our holographic result \eqref{Hol_defect_entropy} is consistent with the UV divergent structure of the defect entropy given in \eqref{defect_entropy_divergence}.
To this end, we regularize the infinite volume of $\BH^{p-1}$ by introducing the small cutoff $\epsilon$ near the boundary $\partial \BH^{p-1}$, resulting in \cite{Graham:1999pm}
\begin{align}\label{regularized_hyp_volume}
    \begin{aligned}
        \text{Vol}_{\epsilon}(\BH^{p-1})
            &=
                \frac{C_{p-2}}{\epsilon^{p-2}} + \frac{C_{p-4}}{\epsilon^{p-4}} +
                \cdots +
                \begin{cases}
                    A\,\log\,\epsilon + \cdots \quad &(p:\text{even}) \ ,\\
                     C_0 \quad &(p:\text{odd}) \ .
                \end{cases}     
    \end{aligned}  
\end{align}
By identifying $\epsilon$ with the UV cutoff in the dual DCFT, we find that \eqref{Hol_defect_entropy} reproduces the general structure of the defect entropy \eqref{defect_entropy_divergence}.
While $C_{p-2}, C_{p-4}, \cdots$ depends on the choice of the cutoff $\epsilon$, the coefficients $A$ and $C_0$ are free from such an ambiguity.
These coefficients can be read off from the renormalized volume of $\BH^{p-1}$:\footnote{This is obtained by the dimensional regularization.}
\begin{align}\label{renormalized_hyp_volume}
        \text{Vol}(\BH^{p-1})
        =
            \frac{\pi^\frac{p}{2}}{\sin\left(\frac{\pi\,p}{2}\right)\,\Gamma\left(\frac{p}{2}\right)} \ ,
\end{align}
whose pole at $p$ even corresponds to the logarithmic divergence in \eqref{regularized_hyp_volume}.

\subsubsection{Holographic defect free energy}\label{subsec:Hol_free_energy}

We now compute the defect free energy \eqref{defect_free_energy} in our model.
The GKPW relation \cite{Gubser:1998bc,Witten:1988hf}  relates the Euclidean on-shell action $I$ of the bulk AdS spacetime and the partition function $Z$ of the dual theory by $I = -\log Z$.
Thus, the defect free energy is written as
\begin{align}\label{defect_free_energy_bulk_DCFT}
    \log\,\langle\,\CD^{(p)}\,\rangle
            &=
                -I(\rho_{\ast},T) + I_{\text{AdS}} \ ,
\end{align}
where $I(\rho_{\ast},T)$ is the action \eqref{HDCFT_action} evaluated on the bulk geometry with Dirichlet boundary condition on $Q$, and $I_\text{AdS}$ is the action of the whole AdS space dual to the ambient CFT.

To calculate the on-shell action, we start with the global AdS coordinates:
\begin{align}\label{AdS_global}
    \d s^2
        =
            L^2\left[
                \d u^2
                +
                \sinh^2u
                \,(\d\varphi^2 + \cos^2\varphi\,\d\Omega_{p}^2 + \sin^2\varphi\,\d\Omega_{d-p-1}^2)\,
                \right] \ .
\end{align}
The dual DCFT lives on $\BS^{d}$ at the boundary of the AdS space $u = \infty$, where the defect wraps on the $\BS^{p}$ at $\varphi = 0$ as in figure \ref{fig:sphere}.
To make contact with the bulk geometry with the EoW brane, we perform the coordinate transformation
\begin{align}
    \cot\varphi = \coth\rho\,\sinh w \ ,
    \quad
    \cosh u = \cosh\rho\, \cosh w \ ,    
\end{align}
to obtain the hyperbolic slicing coordinates:
\begin{align}
    \d s^2
        =
            L^2\left[ 
                \d \rho^2 
                +
                \cosh^2\rho\,(\d w^2 + \sinh^2w\,\d\Omega_{p}^2)\,
                +
                \sinh^2\rho\,\d\Omega_{d-p-1}^2
                \right] \ .
\end{align}
In our model, the brane terminates the geometry at $\rho = \rho_\ast$, and the defect in the dual DCFT wraps a $p$-sphere $\BS^{p}$ at $w = \infty$ at the AdS boundary $\rho=\infty$.

The on-shell action can be calculated either in the global coordinates or the hyperbolic slicing as they give the same result (up to divergences) in accordance with the relation \eqref{partition_function} in the dual theory.
To ease the calculation, we will use the hyperbolic slicing, where the on-shell action \eqref{HDCFT_action} becomes
\begin{align}\label{on-shell}
    \begin{aligned}
        I(\rho_{\ast},T)
            &=
                \frac{L^{d-1}}{8\pi\,G_N}\,\text{Vol}(\BH^{p+1})\,\text{Vol}(\BS^{d-p-1}) \\
                &\qquad\times\bigg[\,d\int^{\infty}_{\rho_\ast}\,\d\rho\,\cosh^{p+1}\rho\,
                \sinh^{d-p-1}\rho \\
                &\qquad\quad + \big\{ (p+1)\tanh\rho_\ast + (d-p-1)\coth\rho_\ast + L\,T\big\}\,\cosh^{p+1}\rho_\ast\,
                \sinh^{d-p-1}\rho_\ast\,\bigg]  \ . 
    \end{aligned}
\end{align}
It follows from \eqref{on-shell} and \eqref{defect_free_energy_bulk_DCFT} with $I_\text{AdS}$ given by setting $\rho_\ast = 0$ and $T=0$ in \eqref{on-shell} that the defect free energy becomes\footnote{Even if we introduce the Gibbons-Hawking term for $M$, its contribution to the defect free energy cancels out between $I(\rho_{\ast},T)$ and $I_{\text{AdS}}$.}
\begin{align}\label{hol_defect_free_energy}
    \begin{aligned}
        \log\,\langle\,\CD^{(p)}\,\rangle
            &=  
                \frac{L^{d-1}}{8\pi\,G_N}\,\text{Vol}(\BH^{p+1})\,\text{Vol}(\BS^{d-p-1}) \\
                &\qquad\times\bigg[\,d\int_0^{\rho_\ast}\,\d\rho\,\cosh^{p+1}\rho\,
                \sinh^{d-p-1}\rho \\
                &\qquad\quad -\big\{ (p+1)\tanh\rho_\ast + (d-p-1)\coth\rho_\ast + L\,T\big\}\,\cosh^{p+1}\rho_\ast\,
                \sinh^{d-p-1}\rho_\ast\,\bigg]  \ . 
    \end{aligned}    
\end{align}
Using the regularized hyperbolic volume \eqref{regularized_hyp_volume}, one can verify that our holographic calculation \eqref{hol_defect_free_energy} reproduces the UV divergent structure \eqref{defect_energy_divergence} of the defect free energy.

Let us examine the relation \eqref{universal} between the defect free energy and the defect entropy in our model.
It follows from \eqref{Hol_defect_entropy} and \eqref{hol_defect_free_energy} that they are related by
\begin{align}\label{S_D_hol_relation}
    \begin{aligned}
         S_{\text{defect}}
            =
                \log\,\langle\,\CD^{(p)}\,\rangle
                &-\frac{L^{d-1}}{4\,G_N}\frac{\pi^\frac{d}{2}}{\sin\left(\frac{\pi\, p}{2}\right)\,\Gamma\left( \frac{p+2}{2}\right)\,\Gamma\left( \frac{d-p}{2}\right)}\,\cosh^{p+1}\rho_\ast\,\sinh^{d-p-1}\rho_\ast
            \\
            &\qquad\times
                \big\{\,p\,\tanh\rho_\ast + (d-p-1)\coth\rho_\ast +L\,T\,\big\} \ .
    \end{aligned}
\end{align}
By comparing with the relation \eqref{universal}, we read off the coefficient $a_T$ of the one-point function $\langle \, T_{\mu\nu}\, \rangle$ of the ambient stress tensor \eqref{one_point_stress} in the dual DCFT:
\begin{align}\label{hol_a_T}
    a_T
        =
            \frac{L^{d-1}}{8\pi\,G_N}\, \frac{d}{d-p-1}\,\left(p\,\tanh\rho_\ast + (d-p-1)\coth\rho_\ast +L\,T\right)\,   
            \cosh^{p+1}\rho_\ast\,\sinh^{d-p-1}\rho_\ast \ .
\end{align}
Since $\rho_\ast$ and $T$ are non-negative for physical solutions of the brane, the coefficient $a_T$ is non-negative in our holographic model.
This is in accordance with the conjecture stating $a_T\ge0$ in unitary DCFTs \cite{Lemos:2017vnx}.

\subsection{Holographic defect $\CC$-theorem}\label{subsec:Hol_RG}

The defect $\CC$-function \eqref{defect_C_function} can be derived straightforwardly from the defect free energy \eqref{hol_defect_free_energy} in our model.

For $p< d-1$, it follows from  \eqref{hol_defect_free_energy} and \eqref{defect_C_function} that the defect $\CC$-function takes the form:
\begin{align}\label{defect_free_energy_univ_DBC}
    \begin{aligned}
        \tilde{D}\,(\rho_\ast,T)
            &=
                \frac{L^{d-1}}{4\,G_N}\,
                \frac{\,\pi^\frac{d}{2}}{\Gamma\left(\frac{d-p}{2}\right)\,\Gamma\left(\frac{p+2}{2}\right)} \\
                    &\qquad\times\bigg[\,-\frac{d}{d-p}\,\tanh^{d-p}\rho_\ast\cdot{}_2F_1 \left(\frac{d-p}{2}, \frac{d+2}{2}, \frac{d-p+2}{2}; \tanh^2\rho_\ast \right) \\
                    &\qquad\qquad
                    +\big\{\, (p+1)\tanh\rho_\ast + (d-p-1)\coth\rho_\ast + L\,T\,\big\}\,\cosh^{p+1}\rho_\ast\,
                    \sinh^{d-p-1}\rho_\ast\,\bigg] \ .
    \end{aligned}
\end{align}
Since $\rho_\ast >0$, the defect $\CC$-function monotonically increases as $T$ increases.
While it is not obvious from \eqref{defect_free_energy_univ_DBC}, one can also show that it is always positive for any $T>0$ as follows.
Since \eqref{defect_free_energy_univ_DBC} is a monotonically increasing function with respect to $T$, we just confirm the positivity at $T=0$.
In this case, we use the integral representations of the hypergeometric function in $\tilde{D}\,(\rho_\ast,T=0)$ and the second term inside the square bracket to obtain
\begin{align}
    \begin{aligned}
        \tilde{D}\,(\rho_\ast,T=0)
            &=
                \frac{L^{d-1}}{4\,G_N}\,
                \frac{\,\pi^\frac{d}{2}}{\Gamma\left(\frac{d-p}{2}\right)\,\Gamma\left(\frac{p+2}{2}\right)} \\
            &\qquad\times
            \bigg[
            \int_0^{\rho_\ast}\,\d\rho\,
            \bigg\{\,
            2\,(p+1)(d-p-1)\cosh^{p+1}\rho\,\sinh^{d-p-1}\rho \\
            &\qquad+p(p+1)\cosh^{p-1}\rho\,\sinh^{d-p+1}\rho+(d-p-1)(d-p-2)\cosh^{p+3}\rho\,\sinh^{d-p-3}\rho
            \,\bigg\} \\ 
            &\qquad+\delta_{p,d-2}
            \bigg] \ , 
    \end{aligned}
\end{align}
where $\delta_{p,d-2}$ comes from the fact that the second term inside the brace of \eqref{defect_free_energy_univ_DBC} has no
$\sinh\rho_{\ast}$ factor when $p=d-2$.
The integrand is manifestly positive, which proves $\tilde{D}\,(\rho_\ast,T)\ge0$ for any $T\ge0$.
We observe that $\tilde{D}\,(\rho_\ast=0,T=0) = 0$ except for $p=d-2$ where $\tilde{D}\,(\rho_\ast=0,T=0) > 0$ due to the last term.

\begin{figure}[t]
    \centering
    \begin{tikzpicture}
        \draw[-{Stealth[angle'=60,scale=0.8]}] (0,0) -- (0,3.0);
        \draw[-{Stealth[angle'=60,scale=0.8]}] (0,0) -- (3.0,0);
        \draw[-{Stealth[angle'=60,scale=0.8]}] (0,0) -- (2.8,2.8);
        \draw[-{Stealth[angle'=60,scale=0.8]}] (0.8,1.7) arc (90:0:1.0 and 1.0);
        \filldraw[orange] (0,0) circle [x radius = 0.05, y radius = 0.05];
        \node at (0,3.3) {$z$};
        \node at (3.3,0) {$r$};
        \node at (3.1,3.1) {$Z$};
        \node at (1.8,0.5) {$\rho$};
        \node at (2.5,1.0) {$N$};
        \node at (1.5,-0.3) {$M$};
        \node at (2.0,2.4) {$Q$};
        \node at (0,-0.3) {$\CD^{(p)}$};
        \node at (-0.4,0.3) {UV};
        \node at (-0.4,2.7) {IR};
    \end{tikzpicture}
    \caption{The relation between the Poincar\'e patch \eqref{Poincare_metric} and the hyperbolic slicing coordinate \eqref{metric_AdS_AdSslice}. The coordinate $Z$ can be viewed as the RG scale.}
    \label{fig:RG}
\end{figure}
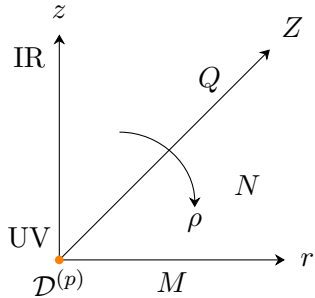

Next, we examine the defect $\CC$-theorem for defect-localized RG flows holographically described by a single real scalar field localized on $Q$ (see also \cite{Kanda:2023zse}):
\begin{align}\label{localized_scalar_model}
    I_Q
        =
            \frac{1}{2}\,\int_Q\d^d y\,\sqrt{h}\left[\, h^{ab}\,\partial_a \phi\, \partial_b\phi + V(\phi)\, \right]\ .
\end{align}
We follow the argument in \cite{Yamaguchi:2002pa} and assume that the potential $V(\phi)$ has some critical points $\phi_0$ satisfying
\begin{align}
    \frac{\d V(\phi)}{\d\phi}\bigg|_{\phi=\phi_0}
        =
            0 \ .
\end{align}
At each critical point $\phi=\phi_0$ (constant), this model reduces to the previous AdS/DCFT model with the tension $T=V(\phi_0)$ and the defect-localized RG flow is triggered by letting $\phi$ roll off from a local maximum to a local minimum of $V(\phi)$.
Let us use the induced metric \eqref{induced_metric} on $Q$ and define
\begin{align}
    T(\phi)
        :=
            V(\phi) - \frac{Z^2}{L^2\cosh^2\rho_\ast}(\partial_Z \phi)^2 \ ,
\end{align} 
which reduces to the brane tension $T=V(\phi_0)=T(\phi_0)$ at each critical point.
Suppose $\phi$ is a function of $Z$; $\phi=\phi(Z)$, then we can show that $T(\phi)$ is a monotonically decreasing function with respect to $Z$ by using the equation of motion of $\phi$:
\begin{align}\label{Monotonicity_T}
    \partial_{Z}T(\phi)
        =
            - \frac{2\,p\,Z}{L^2\cosh^2\rho_\ast}\,(\partial_{Z}\phi)^2
        \leq
            0 \ .
\end{align}
In the Poincar\'e patch \eqref{Poincare_metric}, the $z$ coordinate is identified with the RG scale in the standard AdS/CFT setup.
Since the coordinate $Z$ is linear in $z$, $Z$ can be also regarded as the RG scale in our holographic model (see figure \ref{fig:RG}).
With this identification in mind, the inequality \eqref{Monotonicity_T} implies that the critical value of the potential, i.e., the brane tension $T=V(\phi_0)$ is non-increasing under the RG flow interpolating between the UV
fixed point $\phi_{\text{UV}} = \phi(Z=0)$ and the IR fixed point $\phi_{\text{IR}} = \phi(Z=\infty)$:
\begin{align}
    T_{\text{UV}}\,\geq\,T_{\text{IR}} \ .
\end{align}
Since the defect $\CC$-function \eqref{defect_free_energy_univ_DBC} is monotonic in $T$, we obtain the holographic defect $\CC$-theorem for $p< d-1$:
\begin{align}
    \tilde{D}_{\text{UV}}\,\geq\,\tilde{D}_{\text{IR}} \ .
\end{align}

\section{Correlation functions}\label{sec:Hol_correlators}
We will show that our holographic DCFT model correctly reproduce the correlation functions of the ambient operators in a DCFT.
In section \ref{ss:bul_scalar_EoW}, we will calculate the one-point functions by using a bulk scalar field $\Phi$ coupled to the brane $Q$ \cite{Fujita:2011fp,Kastikainen:2021ybu}.
In section \ref{ss:geodesic_approx}, we employ the geodesic approximation \cite{Kastikainen:2021ybu,Park:2024pkt} for computing the one- and two-point functions.

\subsection{Bulk scalar field coupled to the EoW brane}\label{ss:bul_scalar_EoW}

Let us introduce a bulk scalar field to our holographic model of DCFTs, which coupled to the EoW brane with the action \cite{Fujita:2011fp,Izumi:2022opi}:
\begin{align}\label{action_scalar}
    I_\Phi
        =
            \frac{1}{16\pi\,G_N}\,\int_N \d^{d+1}X\,\sqrt{g}\,\left[\, g^{AB}\,\partial_A \Phi\, \partial_B\Phi + m^2\Phi^2\, \right] 
            -
            \frac{a}{8\pi\,G_N}\,\int_Q \d^d y\,\sqrt{h}\,\Phi \ .
\end{align}
We closely follow \cite{Kastikainen:2021ybu, Almheiri:2018ijj} for computing the one-point function of a scalar operator dual to $\Phi$ below.
To the end, we will neglect the backreaction of the scalar field on the metric and work in the fixed background \eqref{Poincare_metric}.
As in the standard AdS/CFT dictionary, the dual scalar operator has conformal dimension $\Delta$ related to the mass $m$ of the scalar field via $m^2\,L^2 = \Delta\,(\Delta-d)$.
In what follows, we focus on the solution with $\Delta>d/2$.

Taking the variation of the action, we find the Klein-Gordon equation for $\Phi$ while imposing the boundary condition on $Q$:
\begin{align}\label{BC_scalar}
    (z\cos\theta\,\partial_z - z\sin\theta\,\partial_r)\,\Phi\,|_{Q} - a\,L
        =
            0 \ ,
\end{align}
where we define $\tan\theta := \csch\,\rho_{\ast}~(0<\theta\leq\pi/2)$.
We employ the background field technique and decompose the scalar field as 
\begin{align}
    \Phi
        =
            \Phi_B + \tilde{\Phi} \ ,
\end{align}
where the background field $\Phi_B$ is subject to the bulk equation of motion together with the boundary condition \eqref{BC_scalar}.
Moreover, $\Phi_B$ should also be normalizable near the AdS boundary $z \rightarrow 0$:
\begin{align}\label{normalizable}
    \Phi_B
        \sim
            z^{\Delta}\,f_B(x) \ ,
\end{align}
where $x := (\hat{x}^{\hat{a}},x^i_{\perp})$.
In other words, $\Phi_B$ represents the expectation value of the dual operator that arises solely due to the boundary condition (or equivalently the presence of the brane $Q$), even in the absence of an external source.
Then, the fluctuation $\tilde{\Phi}$ is also subject to the Klein-Gordon equation and the boundary condition on $Q$:
\begin{align}\label{BC_fluc}
    (z\cos\theta\,\partial_z - z\sin\theta\,\partial_r)\,\tilde{\Phi}\,|_{Q} 
        =
            0 \ .
\end{align}
In the $z \rightarrow 0$ limit, it behaves as
\begin{align}
    \tilde{\Phi}
        \sim
            z^{d-\Delta}\,J(\hat{x}^{\hat{a}},x^i_{\perp}) + z^{\Delta}\,A(\hat{x}^{\hat{a}},x^i_{\perp}) \ ,
\end{align}
where $A(x)$ is non-locally related  to $J(x)$ by
\begin{align}
    A(x)
        =
            \int\,\d^{d} x'\,J(x')\,H(x,x') \ ,
\end{align}
The kernel $H(x,x')$ is expressed by the bulk-to-boundary propagator $K(z,x;x')$ which satisfies the boundary condition \eqref{BC_fluc} at $Q$ (see \cite{Kastikainen:2021ybu} and references therein for further details):
\begin{align}
    H(x,x')
        =
            \lim_{z \rightarrow 0}\,z^{-\Delta}\,K(z,x;x') \ .
\end{align}

Evaluating the on-shell action of the scalar field \eqref{action_scalar} under these conditions and performing the appropriate renormalization, we find
\begin{align}
    -I_{\Phi,\text{on-shell}}
        =
            \Delta\,\int\,\d^d x\,J(x)\,f_B(x)
            +
            \frac{2\Delta-d}{2}\,\int\,\d^d x\,\int\,\d^d x'\,J(x)\,J(x')\,H(x,x') \ ,  
\end{align}
where we omit the terms that include only $\Phi_B$, since they do not contribute to the correlation functions. 
The scalar one-point function can be obtained by\footnote{This is different from the standard relation given in \cite{Klebanov:1999tb}.
This is because the identification of the dual operator is different due to the existence of the background field $\Phi_B$ \cite{Almheiri:2018ijj}.
}
\begin{align}   
    \langle\,\CO(\hat{x}^{\hat{a}},x^i_{\perp})\,\rangle
        =
            -\frac{\delta\,I_{\Phi,\text{on-shell}}}{\delta\,J(\hat{x}^{\hat{a}},x^i_{\perp})} \bigg|_{J=0}
        =
            \Delta\,f_B(\hat{x}^{\hat{a}},x^i_{\perp}) \ .
\end{align}

We now turn to the calculation of $f_B(x)$.
Defining the following new coordinate
\begin{align}
    \ell    
        :=
            \frac{r}{z}
    \ , \qquad
    v
        :=
            \sqrt{z^2 + r^2} \ ,
\end{align}
the Poincar\'e patch \eqref{Poincare_metric} is transformed to
\begin{align}
    \d s^2
        =
            L^2\,
            \left[\,
            \frac{\d \ell^2}{1+\ell^2} + \frac{1+\ell^2}{v^2}\,(\d v^2 + \d \hat{x}^2_{\hat{a}}) + \ell^2\,\d \Omega^2_{d-p-1}
            \,\right] \ .
\end{align}
Assuming $\Phi_B = \Phi_B(\ell)$, the bulk Klein-Gordon equation of $\Phi_B$ becomes
\begin{align}\label{KG_eq}
    (1+\ell^2)\,\partial^2_{\ell}\,\Phi_B
    +
    \frac{d-p-1+(d+1)\,\ell^2}{\ell}\,\partial_{\ell}\,\Phi_B
    -
    \Delta\,(\Delta-d)\,\Phi_B
    =
    0 \ .
\end{align}
The general solution to this equation for $0<d-p \not\in 2\BZ$ is given by
\begin{align}
    \Phi_B(\ell)
        =
            C_1\,\Phi_1 + C_2\,\Phi_2 \ ,
\end{align}
where $C_{1,2}$ are constants, and $\Phi_{1,2}$ are independent solutions to \eqref{KG_eq}:
\begin{align}
    \Phi_1
        &:=
            \ell^{-d+p+2}\,{}_2F_1 \left(\frac{p-\Delta+2}{2}, \frac{-d+p+\Delta+2}{2}, \frac{-d+p+4}{2}; -\ell^2 \right) \ , 
    \\
    \Phi_2
        &:=
            {}_2F_1 \left(\frac{d-\Delta}{2}, \frac{\Delta}{2}, \frac{d-p}{2}; -\ell^2 \right) \ .
\end{align}
The normalizability \eqref{normalizable} of the background field $\Phi_B$ fixes the ratio of coefficients $C_{1,2}$ as
\begin{align}
    C
        :=
            -\frac{C_2}{C_1}
        =
            \frac{\Gamma(\frac{-d+p+4}{2})\,\Gamma(\frac{\Delta}{2})\,\Gamma(\frac{-p+\Delta}{2})}{\Gamma(\frac{-d+p+\Delta+2}{2})\,\Gamma(\frac{-d+\Delta+2}{2})\,\Gamma(\frac{d-p}{2})} \ .
\end{align}
Furthermore, the boundary condition \eqref{BC_scalar} completely fixes $C_{1,2}$ to be
\begin{align}
    C_1(\theta)
        =
            -\frac{a\,L\,\sin\theta}{F_1(\theta) - C\,F_2(\theta)}
\end{align}
where $F_{1,2}(\theta)$ are defined by
\begin{align}
    \begin{aligned}
        F_1(\theta)
            &:=
                \partial_{\ell}\,\Phi_1\,\big|_{Q}
            \\
            &=
                (-d+p+2)\,(\cot\theta\,)^{-d+p+1}\,\cdot\,{}_2F_1 \left(\frac{p-\Delta+2}{2}, \frac{-d+p+\Delta+2}{2}, \frac{-d+p+2}{2}; -\cot^2\theta \right) \ ,
        \\
        F_2(\theta)
            &:=
                \partial_{\ell}\,\Phi_2\,\big|_{Q}
            =
                \frac{\Delta\,(\Delta-d)}{d-p}\,\cot\theta\,\cdot\,{}_2F_1 \left(\frac{d-\Delta+2}{2}, \frac{\Delta+2}{2}, \frac{d-p+2}{2}; -\cot^2\theta \right) \ . 
    \end{aligned}
\end{align}
From the asymptotic behavior of $\Phi_B$, we read
\begin{align}
    f_B(x)
        =
            \frac{\Gamma(\frac{\Delta}{2})\,\Gamma(\frac{-p+\Delta}{2})}{\Gamma(\frac{d-p-2}{2})\,\Gamma(\frac{-d+2\Delta+2}{2})}\,\frac{C_1(\theta)}{r^{\Delta}} \ .
\end{align}
Finally, the scalar one-point function is obtained as
\begin{align}
    \langle\,\CO(\hat{x}^{\hat{a}},x^i_{\perp})\,\rangle
        =
            \Delta\,f_B(x)
        =
            \frac{2\,\Gamma(\frac{\Delta+2}{2})\,\Gamma(\frac{-p+\Delta}{2})}{\Gamma(\frac{d-p-2}{2})\,\Gamma(\frac{-d+2\Delta+2}{2})}\,\frac{C_1(\theta)}{r^{\Delta}} \ ,
\end{align}
which correctly reproduces the expected form \eqref{one_point_scalar} with the coefficient
\begin{align}\label{a_O_exact}
    a_{\CO}
        =
            \frac{2\,\Gamma(\frac{\Delta+2}{2})\,\Gamma(\frac{-p+\Delta}{2})}{\Gamma(\frac{d-p-2}{2})\,\Gamma(\frac{-d+2\Delta+2}{2})}\,C_1(\theta) \ .
\end{align}

\subsection{Geodesic approximation}\label{ss:geodesic_approx}

We next calculate the one- and two-point functions of scalar operators using the geodesic approximation \cite{Kastikainen:2021ybu,Park:2024pkt}.
While section~\ref{ss:bul_scalar_EoW} was devoted to the analysis of the one-point function, the evaluation of the two-point function is technically more involved. 
Nevertheless, the geodesic approximation provides a practical method for computing both one- and two-point functions for arbitrary $d$ and $p$.

\subsubsection{One-point function}
In the geodesic approximation, the scalar one-point function is given by
\begin{align}
    \langle\,\CO(\hat{x}^{\hat{a}},x^i_{\perp})\,\rangle
        =
            \e^{-\Delta\,L(\hat{x}^{\hat{a}},x^i_{\perp})} \ ,
\end{align}
where $\Delta\,\approx\,mL$ is the conformal dimension of $\CO$, $L(\hat{x}^{\hat{a}},x^i_{\perp})$ the renormalized minimum geodesic length between a boundary point $(\hat{x}^{\hat{a}},x^i_{\perp})$ and a brane point $(\hat{x}^{\hat{a}}_Q,x^i_{Q\perp},z_Q)$ which we will determine later. 
In the Poincar\'e patch \eqref{Poincare_metric}, the geodesic length between a bulk point $(\hat{x}^{\hat{a}}_N, x^i_{N,\perp}, z_N)$ and a boundary point $(\hat{x}^{\hat{a}}_M, x^i_{M,\perp}, 0)$ is given by
\begin{align}\label{bulk_boundary}
    L(\hat{x}^{\hat{a}}_N, x^i_{N,\perp}, z_N;\, \hat{x}^{\hat{a}}_M, x^i_{M,\perp}, 0)
        =
            \log\left[\, 
            \frac{|\hat{x}_N - \hat{x}_M|^2 + |x_{N,\perp} - x_{M,\perp}|^2 + z^2_N}{z_N}
            \,\right] \ .
\end{align}
Then, we can rewrite 
\begin{align}\label{brane_boundary}
    \begin{aligned}
        L(\hat{x}^{\hat{a}},x^i_{\perp})
            &=
                \min
                \bigg[\,
                L(\hat{x}^{\hat{a}}_Q, x^i_{Q,\perp}, z_Q;\, \hat{x}^{\hat{a}}_M, x^i_{M,\perp}, 0)
                \,\bigg] \\
            &=
                \min
                \bigg[\,
                \log\left(\, 
                \frac{|\hat{x}_Q - \hat{x}_M|^2 + |x_{Q,\perp} - x_{M,\perp}|^2 + r^2_Q\tan^2\theta}{r_Q\tan\theta}
                \,\right)
                \,\bigg]
                \ .    
    \end{aligned}
\end{align}
where we used $z_Q = r_Q\,\csch\,\rho_\ast = r_Q \tan\theta$ and the minimization is taken with respect to $\hat{x}^{\hat{a}}_Q$ and $x^i_{Q,\perp}$.
This minimization problem is solved by $\hat{x}^{\hat{a}}_Q = \hat{x}^{\hat{a}}_M,~x^i_{Q,\perp} = x^i_{M,\perp}\cos\theta$, and the result is 
\begin{align} 
    L(\hat{x}^{\hat{a}},x^i_{\perp})
        =
            \log\bigg[2\,|x_{\perp}|\,\tan\frac{\theta}{2}\,\bigg] \ .
\end{align}
Hence, we obtain the expected form of the scalar one-point function
\begin{align}
    \langle\,\CO(\hat{x}^{\hat{a}},x^i_{\perp})\,\rangle
        =
            \frac{1}{2^{\Delta}\tan^{\Delta}\frac{\theta}{2}}\,\frac{1}{|x_{\perp}|^{\Delta}} \ ,
\end{align}
from which we extract the coefficient
\begin{align}\label{a_O_geodesic}
    a_{\CO}
        =
            \frac{1}{2^{\Delta}\tan^{\Delta}\frac{\theta}{2}} \ .
\end{align}

\subsubsection{Two-point function}
A two-point function has contributions from two geodesics which connect two operator insertion points on the boundary $M$: the direct geodesic and the reflecting geodesic.
The direct geodesic length can easily be obtained from \eqref{bulk_boundary} by taking the limit $z_N \rightarrow 0$ with renormalization:
\begin{align}\label{nonreflect}
    L(\hat{x}^{\hat{a}}_1, x^i_{1,\perp};\, \hat{x}^{\hat{a}}_2, x^i_{2,\perp})
        =
            \log\left[\, 
            |\hat{x}_1 - \hat{x}_2|^2 + |x_{1,\perp} - x_{2,\perp}|^2 
            \,\right] \ .
\end{align}
We now compute the geodesic length reflected on the brane $Q$ once.
For simplicity, we only consider the case in which $|x_{1,\perp}| = |x_{2,\perp}| =: r$.
The geodesic length can be written as
\begin{align}
    \begin{aligned}
        L(\hat{x}^{\hat{a}}_1, x^i_{1,\perp};\, \hat{x}^{\hat{a}}_2, x^i_{2,\perp};\, \hat{x}^{\hat{a}}_Q, x^i_{Q,\perp}, z_Q)
            &=
                \log\left[\, 
                \frac{|\hat{x}_1 - \hat{x}_Q|^2 + |x_{1,\perp} - x_{Q,\perp}|^2 + z^2_Q}{z_Q}
                \,\right] \\
            &\qquad +
                \log\left[\, 
                \frac{|\hat{x}_2 - \hat{x}_Q|^2 + |x_{2,\perp} - x_{Q,\perp}|^2 + z^2_Q}{z_Q}
                \,\right] \ ,
    \end{aligned}
\end{align}
where the reflecting point $(\hat{x}^{\hat{a}}_Q, x^i_{Q,\perp}, z_Q)$ on the brane $Q$ will be determined by minimizing this expression.
The variational equations $\partial L/ \partial\hat{x}^{\hat{a}}_Q = 0$ with respect to $\hat{x}^{\hat{a}}_Q$ are satisfied if
\begin{align}\label{variation_sol}
    \hat{x}^{\hat{a}}_Q
        =
            \frac{\hat{x}^{\hat{a}}_1 + \hat{x}^{\hat{a}}_2}{2} \ ,
    \qquad
    |x_{1,\perp} - x_{Q,\perp}|
        =
            |x_{2,\perp} - x_{Q,\perp}| \ .
\end{align}
Under these conditions,\footnote{For a particular range of $\theta$, there exists a solution other than \eqref{variation_sol}.
We assume that \eqref{variation_sol} gives a reflecting geodesic with minimal length in the following calculations. 
} the variational equations $\partial L/ \partial{x}^{i}_{Q,\perp} = 0$ determine ${x}^{i}_{Q,\perp}$.
Here, we record the quantities necessary for the calculation of the geodesic length
\begin{align}
    r_Q
        &=
            \cos\theta\,
            \left( 
            \frac{|\hat{x}_1 - \hat{x}_2|^2}{4} + r^2
            \right)^\frac{1}{2} \ ,
    \\
    x_{1,\perp} \cdot x_{Q,\perp}
        &=
            x_{2,\perp} \cdot x_{Q,\perp}
        =
            r_Q 
            \left(
            \frac{r^2 + (x_{1,\perp} \cdot x_{2,\perp})}{2}
            \right)^\frac{1}{2} \ .    
\end{align}
Then, the reflecting geodesic length is given by
\begin{align}\label{reflect}
    L(\hat{x}^{\hat{a}}_1, x^i_{1,\perp};\, \hat{x}^{\hat{a}}_2, x^i_{2,\perp};\, \hat{x}^{\hat{a}}_Q, x^i_{Q,\perp}, z_Q)
            &=
                2\,\log\left[\, 
                \frac{2\,r}{\sin\theta}
                \left(\,
                \sqrt{\frac{|\hat{x}_1 - \hat{x}_2|^2}{4\,r^2}+1}
                -
                \cos\theta\,\sqrt{\frac{x_{1,\perp} \cdot x_{2,\perp}}{2\,r^2}+\frac{1}{2}}
                \,\right)
                \,\right] \ ,
\end{align}
and the two-point function is obtained by summing two contributions \eqref{nonreflect} and \eqref{reflect}:
\begin{align}
    \langle\,\CO_1(x_1)\,\CO_2(x_2)\,\rangle
        =
            \frac{1}{r^{2\Delta}}
            \bigg[\,
            \xi_1^{-\Delta}
            +
            \left(\frac{\sin\theta}{2}\right)^{2\Delta}
            \bigg(\,            \sqrt{\frac{\xi_1+2\,\xi_2+2}{4}}
            -
            \cos\theta\,\sqrt{\frac{1+\xi_2}{2}}
            \,\bigg)^{-2\Delta}     
            \,\bigg] \ ,
\end{align}
where we used the cross-ratios \eqref{cross-ratios}.
This is in accordance with the generic form of the scalar two-point function in DCFTs \eqref{two_point_scalar}.
For $p=d-1$, it reproduces the result in the AdS/BCFT model \cite{Kastikainen:2021ybu,Park:2024pkt} by setting 
$\xi_2 = 1$.

\section{Discussion}\label{sec:discussion}
In this paper, we have proposed a bottom-up holographic model of DCFTs with defects of codimension greater than one.
The key idea was to treat defects as boundary conditions on a small tubular neighborhood and realize their holographic duals as a limit of the AdS/BCFT model.
Unlike the AdS/BCFT, we employ Dirichlet boundary conditions to allow for consistent solutions with finite bulk regions and positive brane tensions.

We tested our model through several holographic calculations.
The defect entropy and free energy were shown to reproduce the expected UV structures for $p$-dimensional defects.
We also calculated the defect $\CC$-function holographically, showed it is non-negative for physical brane configurations, and proved a holographic defect $\CC$-theorem for defect-localized RG flows triggered by a localized scalar field on the brane.
Moreover, we computed the correlation functions of scalar primaries by two independent methods and verified that both reproduce the expected forms dictated by the symmetry of DCFTs.

The results of this paper give rise to several questions that deserve further investigation.
In contrast to the AdS/BCFT model, the brane tension $T$ and the position $\rho_\ast$ in our model can be chosen as independent parameters.
In the limit with $T\to 0$ and $\rho_\ast \to 0$, our model reduces to the pure AdS space without the EoW brane for $p < d-2$ as seen from the vanishing of the coefficient $a_T$ and the defect $\CC$-function $\tilde D$ given by \eqref{hol_a_T} and \eqref{defect_free_energy_univ_DBC}.
On the other hand, $a_T$ and $\tilde D$ remain finite for $p = d-2$ in this limit, which implies that the resulting solution is locally AdS away from $\rho=0$, but differs from pure AdS.
Although the precise interpretation of the geometry is not clear, this solution could be related to monodromy defects that exist only for $p=d-2$ \cite{Billo:2013jda,Gaiotto:2013nva}. 
It also remains open whether our model can be embedded into string theory.
Such an embedding has been given in the AdS/BCFT model \cite{Fujita:2011fp}, and similar embedding may exist for our setup.

Our model has potential applications and extensions to various directions.
For instance, one may introduce black holes in the bulk and construct a holographic dual of DCFTs at finite temperature.
Such a model would be beneficial to explore the thermal properties and phase structures of DCFTs which remain to be understood.
It would also be interesting to generalize our model to the case with more than one defect such as composite defects \cite{Shimamori:2024yms,Ge:2024hei} or intersecting defects \cite{Shachar:2022fqk}.
To this end, it would be useful to adopt the nested slicing of AdS space by the sub-AdS spaces which has been exploited in constructing Janus-in-Janus geometry dual to nested interfaces \cite{Hirano:2006as} (see also \cite{Gutperle:2020gez}).
Another promising application is the holographic study of conformal blocks in DCFTs.
Such analyses have been already performed for BCFTs \cite{Aharony:2003qf,Karch:2017fuh,Rastelli:2017ecj,Karch:2017wgy,Sato:2017gla}, and our model provides a natural setting to extend them to defects of higher codimension.
We hope that these questions will be addressed in future studies.

\acknowledgments
We are grateful to K.\,Jensen and T.\,Takayanagi for valuable discussions.
The work of T.\,N. was supported in part by the JSPS Grant-in-Aid for Scientific Research (B) No.\,24K00629, Grant-in-Aid for Scientific Research (A) No.\,21H04469, and
Grant-in-Aid for Transformative Research Areas (A) ``Extreme Universe''
No.\,21H05182 and No.\,21H05190.


\bibliographystyle{JHEP}
\bibliography{DCFT}

\end{document}